\documentclass[aps,floats,showpacs,showkeys,preprintnumbers,nofootinbib]{revtex4}
\usepackage{graphicx}

\usepackage{bbm}                                                  

\newcommand{\nc}{\newcommand}
\nc{\beq}{\begin{equation}}  \nc{\eeq}{\end{equation}}
\nc{\bea}{\begin{eqnarray}}  \nc{\eea}{\end{eqnarray}}
\nc{\baa}{\begin{array}}     \nc{\eaa}{\end{array}}
\nc{\bit}{\begin{itemize}}   \nc{\eit}{\end{itemize}}
\nc{\ben}{\begin{enumerate}} \nc{\een}{\end{enumerate}}
\nc{\bce}{\begin{center}}    \nc{\ece}{\end{center}}
\nc{\bpm}{\begin{pmatrix}}   \nc{\epm}{\end{pmatrix}}
\nc{\bvt}{\begin{verbatim}}  \nc{\evt}{\end{verbatim}}
\def\to{\rightarrow}

\def\gesim{\,{\raise-3pt\hbox{$\sim$}}\!\!\!\!\!{\raise2pt\hbox{$>$}}\,}
\def\lesim{\,{\raise-3pt\hbox{$\sim$}}\!\!\!\!\!{\raise2pt\hbox{$<$}}\,}
\def\boldoverdot{\,{\raise6pt\hbox{\bf.}\!\!\!\!\>}}
\def\re{{\bf Re}}
\def\im{{\bf Im}}

\def\acal{{\cal A}}
\def\bcal{{\cal B}}

\def\dcal{{\cal D}}

\def\fcal{{\cal F}}
\def\gcal{{\cal G}}

\def\kcal{{\cal K}}
\def\lcal{{\cal L}}
\def\mcal{{\cal M}}

\def\ocal{{\cal O}}

\def\ucal{{\cal U}}
\def\vcal{{\cal V}}

\def\lll{{\bf l}}

\def\nn{{\bf n}}

\def\xx{{\bf x}}
\def\yy{{\bf y}}
\def\zz{{\bf z}}
\def\aBB{{\mathbbm A}}

\def\mBB{{\mathbbm M}}
\usepackage{bm}
\def\alpbf{{\bm\alpha}}         
\def\sigbf{{\bm\sigma}}         

\def\vev{vacuum expectation value}
\def\irrep{irreducible representation}

\def\rhs{right hand side\ }

\def\Vev{Vacuum expectation value}

\def\det{\hbox{det}}

\def\diag{\hbox{\diag}}

\def\vevof#1{\left\langle#1\right\rangle}
\def\doubleundertext#1{
{\undertext{\vphantom{y}#1}}\par\nobreak\vskip-\the\baselineskip\vskip4pt%
\undertext{\hbox to 2in{}}}
\def\inbox#1{\vbox{\hrule\hbox{\vrule\kern5pt
     \vbox{\kern5pt#1\kern5pt}\kern5pt\vrule}\hrule}}
\def\sqr#1#2{{\vcenter{\hrule height.#2pt
      \hbox{\vrule width.#2pt height#1pt \kern#1pt
         \vrule width.#2pt}
      \hrule height.#2pt}}}
\def\square{\mathchoice\sqr56\sqr56\sqr{2.1}3\sqr{1.5}3}
\def\today{\ifcase\month\or
  January\or February\or March\or April\or May\or June\or
  July\or August\or September\or October\or November\or December\fi
  \space\number\day, \number\year}
\def\pmb#1{\setbox0=\hbox{#1}%
  \kern-.025em\copy0\kern-\wd0
  \kern.05em\copy0\kern-\wd0
  \kern-.025em\raise.0433em\box0 }
\def\up#1{^{\left( #1 \right) }}
\def\inv#1{\frac1{#1}}
\def\su#1{{SU(#1)}}
\def\ui{U(1)}
\def\sumprime_#1{\setbox0=\hbox{$\scriptstyle{#1}$}
  \setbox2=\hbox{$\displaystyle{\sum}$}
  \setbox4=\hbox{${}'\mathsurround=0pt$}
  \dimen0=.5\wd0 \advance\dimen0 by-.5\wd2
  \ifdim\dimen0>0pt
  \ifdim\dimen0>\wd4 \kern\wd4 \else\kern\dimen0\fi\fi
\mathop{{\sum}'}_{\kern-\wd4 #1}}
\newcounter{problem}[section]

\global\advance\voffset by 1 in

\nc{\non}{\nonumber}
\nc{\hc}{\hbox {h.c.}} 
\def\lsim{\mathrel{\raise.3ex\hbox{$<$\kern-.75em\lower1ex\hbox{$\sim$}}}}
\def\gsim{\mathrel{\raise.3ex\hbox{$>$\kern-.75em\lower1ex\hbox{$\sim$}}}}

\nc{\Lsp}{\;\;\;\;\;\;\;\;\;\;}  
\nc{\LLLsp}{\lspace \lspace}
\nc{\lsp}{\;\;\;\;\;\;}
\nc{\spac}{\;\;\;}
\nc{\noi}{\noindent}
\nc{\lra}{\longrightarrow}

\def\scalar{\chi} 
\def\e5{e_5}

\def\mati{{\mathbbm1}}
\def\diag{\hbox{diag}}

\def\vevof#1{{\left\langle#1\right\rangle}}
\nc{\avev}{\vevof{A_4}}
\nc{\eps}{\epsilon}
\nc{\dcp}{\delta^{(CP)}_n}

\begin{document}

\preprint{IFT-05-01\cr
UCRHEP-T384}

\title{Majorana Fermions and CP Violation \\ from 5-dimensional Theories; a Systematic Approach}

\author{Bohdan GRZADKOWSKI}
\email{bohdan.grzadkowski@fuw.edu.pl}
\affiliation{Institute of Theoretical Physics,  Warsaw University,
Ho\.za 69, PL-00-681 Warsaw, Poland}

\author{Jos\'e WUDKA}
\email{jose.wudka@ucr.edu}
\affiliation{Department of Physics, University of California,
Riverside CA 92521-0413, USA}

\begin{abstract}
Within five-dimensional compactified theories we discuss generalized periodicity and orbifold boundary conditions 
that allow for mixing between particles and anti-particles after a shift by the size of extra dimensions or after 
the orbifold reflection. A systematic strategy for constructing 4-dimensional models is presented, 
in particular we find a general form of the periodicity and orbifold conditions that are allowed by consistency 
requirements. We formulate general conditions for a presence of massless Kaluza-Klein modes and discuss remaining 
gauge symmetry of the zero-mode sector. It is shown that if the orbifold twist operation transforms particles 
into anti-particles then the zero-mode fermions
are 4-dimensional Majorana fermions. The possibility of explicit and spontaneous CP violation is discussed. 
General considerations are illustrated by many  Abelian and non-Abelian examples.
\end{abstract}

\pacs{11.10.Kk, 11.30.Er, 11.30.Qc, 12.20.Ds}
\keywords{gauge theories, extra dimensions, charge conjugation, 
           boundary conditions, CP violation}

\maketitle
\section{Introduction}

In the Standard Model (SM), the Higgs mechanism is responsible for generation of 
fermion and vector-boson masses. 
This mechanism, though it leads to renormalizable
and unitary theories, has severe naturality problems
associated with the so-called ``hierarchy problem''~\cite{Georgi:1974yf}.
The tree-level version of this problem reduces to the fact that a possible
(and in the context of grand unified theories even necessary) huge ratio
of mass scales is adopted without any explanation (aside from a desire to make
these models phenomenologically viable). Radiative
corrections usually exacerbate this problem as the quadratic
corrections to the scalar masses tend to destabilize the
original ratio, which requires order-by-order fine tuning
of the parameters.

Extra dimensional extensions of the SM offer a novel
approach to the gauge symmetry breaking in which the hierarchy problem could be
either solved or at least reformulated in terms of geometry of the higher-dimensional space.
Among various attempts in this direction it is worth mentioning the following:
\bit
\item 
The spontaneous breaking of gauge symmetries by imposing non-trivial 
boundary conditions along the extra (compactified) dimensions; the so called 
Scherk-Schwarz (SS) mechanism~\cite{Cremmer:1979uq}.
\item
Symmetry breaking through a non-zero vacuum expectation value of
extra components of the higher dimensional gauge fields; the Hosotani 
mechanism~\cite{Hosotani:1983xw}~\footnote{If the symmetry to be broken is a local one, then
the Hosotani mechanism is equivalent to the Scherk-Schwarz breaking, see e.g. \cite{Quiros:2003gg}.}.
\item
Gauge symmetry braking by asymmetric boundary conditions (BC)
in models of extra dimensions compactified on an interval~\cite{Csaki:2003dt}.
\eit
It is worth noting that even though 5D gauge theories are non-renormalizable, nevertheless,
as is has been recently verified~\cite{SekharChivukula:2001hz}, 
the effective 4-dimensional (4D) theories are tree-level unitary.

In a recent publication~\cite{Grzadkowski:2004jv} we have shown that 5D Quantum 
Electrodynamics compactified on a circle
violates CP either explicitly through a non-symmetric BC or, what is  theoretically much more
appealing, spontaneously through a non-zero one-loop vacuum expectation value for the
zero Kaluza-Klein (KK) mode of the extra component of the U(1) gauge field. The
implementation of this idea in a
realistic model of CP violation (CPV) based on a 5D theory requires that the theory
produce the correct chiral and flavor structures in the light sector. We consider
this latter issue in this paper.

In order to produce a chiral
effective 4D theory we will follow a standard approach and consider 
a 5-dimensional gauge theory compactified on the
 $S_1/Z_2$ orbifold. We will, however, modify and generalize the usual treatment by
allowing non-standard twist operations. Specifically ($y$ denotes the coordinate of
$S_1/Z_2$ and $L$ the radius of $S_1$) under the translation
$y\to y+L$, or under orbifold $Z_2$ reflection $y\to -y$ we will
allow mixing between particles and their charge-conjugated counterparts. 
Such mixing offers particularly useful way to construct models that generate spontaneous CPV 
in the same spirit as in~\cite{Grzadkowski:2004jv}; theories of this type are characterized 
by a non-standard orbifold parity for the fifth component of the gauge field, as only then the
corresponding zero mode survives, and it is the \vev\ of this zero mode that is responsible
for CPV. In this case, however the corresponding 4D components do {\em not} have a zero mode,
and this corresponds to a reduction of the light-sector gauge group. We show below that this
situation is indeed realized when non-trivial orbifold boundary conditions are chosen.

Allowing the boundary conditions to mix particles and anti-particles often reduces by
one half the fermionic degrees of freedom and the surviving 
KK modes (including zero modes) behave as 4D Majorana fermions. 
Such a mechanism will be described below and will be of use when constructing models for
neutrino physics within the context of higher dimensional theories

The paper is organized as follows. In Section~\ref{general} we fix our notation and
we consider the basic properties of the 5D theory including gauge symmetry and discrete
symmetries.
Section~\ref{lightfer} contains discussion of zero mode sector with general conditions which
must be satisfied for the existence of zero modes.
In Section~\ref{sect:abelian.case} we illustrate the general discussion within
Abelian theories containing one or two fermionic fields.
Section~\ref{sect:nonabelian.case} shows non-Abelian examples of models
with the generalized BC.
Summary and conclusions are presented in Section~\ref{summary}.
The appendix contains detailed discussion of the single Abelian fermion and of the
possibility for the spontaneous CPV.

\section{General considerations}
\label{general}

\subsection{The Lagrangian}
We will consider a general 5-dimensional (5D) gauge theory with the gauge fields $A_M$
coupled to a fermionic multiplet $\Psi$. The corresponding Lagrangian takes the form
\beq
\lcal = -\inv4 \sum_a \inv{g_a^2} \; F^a_{MN}F^{a\;MN} + \bar\Psi(i\gamma^N D_N - M ) \Psi \,,
\label{lag}
\eeq
where $D_N = \partial_N + i g_5 A_N$, $A_N = A_N^a T^a$.
We assume a general gauge group (not necessarily simple),
where the gauge couplings are all expressed in units of $g_5$
(which proves convenient since these couplings are not
dimensionless) and are absorbed in the definition of the
gauge fields; the group generators $T_a$ are assumed to
be Hermitian.
All fermions are collected in the multiplet $ \Psi $ that
is in general reducible, and may contain
several sub-multiplets transforming according
to the same gauge-group \irrep\footnote{We could, in principle, consider
a non-standard kinetic term, $\bar\psi Z \gamma^N\partial_N \psi$, 
with $Z$  a Hermitian matrix. However, if we restrict 
ourselves to non-tachyonic theories, then the eigenvalues
of $Z$ must be positive, so in the diagonal basis we have 
$Z=K^2$, where $K$ is a real diagonal matrix. The rescaling 
$\psi\to K^{-1}\psi$ would bring the kinetic term to 
its standard canonical form adopted in (\ref{lag}).}.  
We will allow all fields to propagate throughout the
5D manifold.

We assume that the global topology of the 5-dimensional space-time
is $ \mBB^4 \times (S_1/Z_2) $. We denote by $ x^\mu, ~\mu=0,\ldots,4$
the $\mBB^4$ coordinates (with $x^0$ the only time-like direction); 
and by $y$ that of $ S_1/Z_2$, with 
$ 0 \le y \le L $ and $y$ identified with $-y$.
The metric is assumed to be flat, with convention 
$g_{NM}={\rm diag}(1,-1,-1,-1,-1)$ with the last entry associated with $S_1/Z_2$.

Given this space-time structure the fields can have
non-trivial boundary conditions (BC) in the $y$ coordinate, both
under translations $y \to y+L$ and inversions $y\to-y$. 
We will consider the most general boundary conditions allowed by
the gauge and Lorentz (in $ \mBB^4$) symmetries, which may
involve both the fields and their
charge conjugates. Such mixed BC can violate
some of the global symmetries, and the conditions under which this occurs
and its consequences will be investigated below.
It is worth emphasizing that the theory is defined by (\ref{lag})
together with the imposed BC, all of which we assume fixed. 
A transformation that alters the BC maps one
theory to another and cannot be understood to be a symmetry of
the original theory.

In order to simplify our notation we will often
suppress the dependence of fields on $x$,  
and write e.g. $\chi(y)$ or $A_M(y)$ instead of
$\chi(x,y)$ and $A_M(x,y)$, respectively; it should be understood that 
whenever a field depends on $y$ it is also a function of $x$.

\subsection{Periodicity}

As mentioned above we will discuss generalized periodicity 
conditions that allow mixing between particles and 
anti-particles:
\bea
\Psi_{L/R}(y+L) &=& \Gamma_{L/R} \Psi_{L/R}(y) + 
    \Upsilon^*_{L/R} \Psi^c_{L/R}(y) \cr
A_N(y+L) &=& \left\{ \baa{lr}
                       +U_1^\dagger A_N(y)   U_1 & (P1) \\ 
                       -U_2^\dagger A_N^T(y) U_2 & (P2) 
                     \eaa
\right.\,,
\label{Pbc}
\eea
where $\Psi^c$ denotes the charged-conjugate field  
$ C_5\left(\bar\psi\right)^T$ with
$C_5$ the 5D charge conjugation operator defined by the relation 
$ \gamma_0^* C_5^\dagger \gamma_0 \gamma_N C_5 = \gamma_N^T $~\footnote{
Whenever an explicit representation is needed for the Dirac 
matrices we will adopt the Dirac representation. In this case
$C_5=\gamma_1\gamma_3$. The 5D parity, which will be relevant later, 
is defined by $\Psi\to P\Psi$ with $P=\gamma_0\gamma_4$.
We also choose $i \gamma^4 = + \gamma_5 =i\gamma^0\gamma^1\gamma^2\gamma^3$.
Note that in 5D, the parity reflection is defined~\cite{BelenGavela:1983tp} such that one spatial
component is preserved: $x^{0,4}\to x^{0,4}$ and $x^i\to -x^i$, for $i=1,2,3$.},
$U_{1,2}$ are global elements of the gauge group while the
matrices $\Gamma$ and 
$\Upsilon$ are matrices constrained by requiring invariance of 
$ \lcal$ under
the so called twist operation defined by the \rhs of (\ref{Pbc}).
In particular, 
it is easy to see that the  invariance of the 
fermionic kinetic term $\bar\Psi i\gamma^N \partial_N \Psi$
requires $ \Gamma_L = \Gamma_R = \Gamma $ and $ \Upsilon_L = 
\Upsilon_R = \Upsilon $. Therefore we
will consider only the non-chiral BC
$\Psi(y+L) = \Gamma \Psi(y) + \Upsilon^* \Psi^c(y)$.
Note that the matrices $ \Gamma$ and $ \Upsilon$ in general affect
both flavor and gauge indices.

The  motivation for considering the option P2 is to allow
for the presence of the charge-conjugate gauge fields in the BC
in parallel with our choice of fermionic boundary conditions, which
also involve charge-conjugate fields.
It should be emphasized that a linear combination of P1 and P2 
is not allowed since it does not leave the gauge kinetic term 
invariant\footnote{For Abelian groups there appears to be an 
additional sign freedom since the gauge-kinetic term is even in 
$A$. One can verify, however, that this is in fact covered
by either P1 or P2.}.

In describing the constraints imposed by the invariance of $ \lcal$ under
(\ref{Pbc}) it proves convenient to introduce the following notation 
\beq
\begin{array}{lll}
\chi \equiv \left( \begin{array}{c} \Psi^c \cr \Psi \end{array} \right) &
\acal \equiv \left( \begin{array}{cc}
                  \Gamma & -\Upsilon^* \cr 
                  \Upsilon & \Gamma^* 
                 \end{array} \right) & \cr
\tau^a  \equiv \left( \begin{array}{cc} 
                    T_a & 0 \cr 
                    0 & - T_a^* 
                 \end{array} \right) &
\ucal_1 \equiv \left( \begin{array}{cc} U_1 & 0 \cr 0 & U_1^* \end{array} \right) &
\ucal_2 \equiv \left( \begin{array}{cc} 0 & U_2^* \cr U_2 & 0 \end{array} \right)
\end{array}
\label{adef}
\eeq
in terms of which the fermionic periodicity conditions are simply
\beq
\chi(y+L) = \acal^* \chi(y)\,.
\label{period}
\eeq
Requiring invariance of the kinetic term $\bar\Psi i\gamma^N D_N \Psi$ 
gives the following conditions on the acceptable BC:
\beq
\acal^\dagger \acal = 1, \quad
\baa{l}
P1:\quad [\tau_a , \ucal_1 \acal] = 0 \\
P2:\quad [\tau_a , \ucal_2 \acal] = 0 
\eaa
\,.
\label{pcon}
\eeq
Decomposing $ \Psi $ into a 
set of gauge multiplets $ \{\psi_r\}$ each transforming as an \irrep\
of the gauge group,  we find that $ \acal$
can mix $ \psi_r $ with $ \psi_s$ (via $ \Gamma$) 
or with $ \psi^c_u$ (via $ \Upsilon$) {\em provided}
$ \psi_s $ and $ \psi^c_u$ belong to the same \irrep\ as $ \psi_r $.
We will discuss this in detail in section \ref{gen_sol}. 

The conditions for the mass term to be invariant under the twist operation can be derived
in the same way, we find 
\beq
[\acal,\mcal] = 0 ; \qquad
\mcal \equiv \left( \begin{array}{cc} M & 0 \cr 0 & -M^* \end{array} \right)\,.
\eeq

We note that it is possible to choose a basis where the fermion
fields are simply periodic: writing $\acal^*=e^{i\kcal}$, the
fields
(see also \cite{Biggio:2002rb}):
\beq
\chi\,'(y)=e^{-i\kcal y/L}\chi
\eeq
satisfy $ \chi\,'(y+L)=\chi\,'(y)$.
Such a transformation, however, generates a non-standard $y$
and $\acal$-dependent mass term. We do not use the above
field redefinition because of this complication.

\subsection{Orbifold reflections}

In a similar way we adopt the most general twist transformation for the
orbifold reflection $ y \to - y $. The BC read:
\bea
\chi(-y) &=& \gamma_5 \bcal^* \chi(y) \cr
A_N(-y) &=& 
\left\{
\baa{lr}
 (-1)^{s_N}   \tilde U_1^\dagger A_N(y)   \tilde U_1 & (R1)\\
 (-1)^{1-s_N} \tilde U_2^\dagger A_N^T(y) \tilde U_2 & (R2)
\eaa
\right.\,,
\label{Rbc}
\eea
where $ s_N= \delta_{N,4} $, $ \tilde U_{1,2} $  are global 
gauge transformations and
\beq
\bcal\equiv
\left(
\baa{cc}
-\tilde\Gamma & \tilde\Upsilon^*\\
\tilde\Upsilon & \tilde\Gamma^*
\eaa
\right)\,.
\label{bdef}
\eeq
Requiring now the invariance of $ \lcal $ under (\ref{Rbc})
implies
\beq
\bcal^\dagger \bcal = 1\,, \quad
\baa{ll}
R1:& [ \tau_a , \tilde\ucal_1 \bcal ] = 0 \cr
R2:& [ \tau_a , \tilde\ucal_2 \bcal ] = 0 
\eaa
\,,
\label{rcon}
\eeq
where
\beq
\tilde\ucal_1 \equiv  \left( \begin{array}{cc}
                          \tilde U_1 & 0 \cr 
                          0 & \tilde U_1^* 
                       \end{array} \right) 
\qquad
\tilde\ucal_2 \equiv  \left( \begin{array}{cc}
                          0 & \tilde U_2^* \cr 
                          \tilde U_2 & 0
                       \end{array} \right) 
\,.
\eeq

The mass term is invariant under the orbifold twist (\ref{Rbc}) provided
\beq
\{\bcal,\mcal\}= 0 \,.
\eeq

\subsection{Consistency conditions}

The periodicity and reflection transformations are not independent
since $ - y = [-(y+L)]+L $ and $ -(-y) = y $. These imply,
respectively,
\bea
\bcal & =& \acal \bcal \acal\,, \label{baba.f}\\
\bcal^2 &=& \mati \label{bb.f}
\eea
for the fermions. For the $Pi-Rj$ BC ($i,j=1,2$) the corresponding
constraints on the gauge bosons give (no sum over $i$ and $j$)
\bea
&& [\tau_a, \tilde\vcal_j \vcal_i \tilde\vcal_j \vcal_i^\dagger ] = 0 \label{baba.b} \,,\\
&& [\tau_a, \tilde\vcal_i^2 ] = 0 \,,\label{bb.b}
\eea
where $ \vcal_1 = \ucal_1,~ \tilde\vcal_1 = \tilde\ucal_1$
and $ \vcal_2 = \ucal_2^*,~ \tilde\vcal_2 = \tilde\ucal_2^*$.

These  conditions imply that $\tilde\vcal_j \vcal_i \tilde\vcal_j 
\vcal_i^\dagger$ and $ \tilde\vcal_i^2 $ belong to the center of
the group. If the representation generated by $\{ \tau_a \}$
is split into its irreducible components, the projection of 
these matrices onto each irreducible subspace
must be proportional to the unit matrix as a consequence of the Schur's lemma. We now examine
this and other similar restrictions imposed by the local symmetry.

\subsection{Gauge invariance}
\label{sect:gi}

Under a gauge transformation $ \Omega $ the fields transform as
\beq
A_N \to A_N' = \inv{i g_5} \Omega D_N \Omega^\dagger \qquad {\rm and} \qquad
\chi \to \chi' = \left( \begin{array}{cc}
                            \Omega^* & 0 \cr
                            0 & \Omega
                        \end{array} \right)\chi \equiv \ocal^* \chi\,.
\label{gauge.transf}
\eeq
For the theory to be gauge invariant the gauge-transformed fields should satisfy 
the same boundary conditions (\ref{Pbc}) and (\ref{Rbc}):
\bea
\chi'(y+L) = \acal^* \chi'(y)\qquad\qquad\qquad\qquad &  
\chi'(-y) = \gamma_5 \bcal^* \chi'(y) \label{f_gi}\qquad\qquad\qquad\qquad\quad\\
A'_N(y+L) = 
\left\{ \baa{lr}
                       +U_1{}^\dagger A'_N(y)     U_1 & (P1)\cr
                       -U_2{}^\dagger A'_N{}^T(y) U_2 & (P2)  
        \eaa
\right. & 
A'_N(-y) = 
\left\{ \baa{lr}
    (-1)^{s_N}   \tilde U_1^\dagger A'_N(y)  \tilde U_1 & (R1)\cr
    (-1)^{1-s_N} \tilde U_2^\dagger A_N^T(y) \tilde U_2 & (R2)
        \eaa
\right. \,. 
\label{b_gi}
\eea

We consider first the constraints implied by imposing
P1. Using the transformation properties of $ \chi$ we
find that this choice of BC respects gauge invariance provided
\beq
\ocal(y+L) = \acal \ocal(y) \acal^\dagger \qquad (P1)\,.
\label{chi.gi}
\eeq
Similarly, the transformation properties of the gauge fields require
\beq
\left(\Omega D_N \Omega^\dagger \right)_{y+L} = \left(U_1^\dagger 
\Omega D_N \Omega^\dagger U_1 \right)_y \,,
\eeq
which leads to
\bea
&& \left(\Omega \partial_N \Omega^\dagger \right)_{y+L} = \left(U_1^\dagger 
\Omega \left(\partial_N \Omega^\dagger\right) U_1 \right)_y \,,\cr
&& \left[ T_a , \Omega^\dagger(y) U_1 \Omega(y+L) U_1^\dagger \right] = 0\,,
\eea
where we used (\ref{Pbc}) to express $A_N(y+L)$ in terms
of $A_N(y) $. In terms of $ \ocal$ these constraints become
\bea
&& \left(\ocal \partial_N \ocal^\dagger \right)_{y+L} = \left(\ucal_1^\dagger 
\ocal  \left(\partial_N \ocal^\dagger\right) \ucal_1 \right)_y \,,\cr
&& \left[ \tau_a , \ocal^\dagger(y) \ucal_1 \ocal(y+L) \ucal_1^\dagger \right] = 0 \,.
\eea
Using then (\ref{chi.gi}) we find
\bea
&& \left[\ocal \partial_N \ocal^\dagger , \ucal_1 \acal \right] = 0 \,,\cr
&& \left[ \tau_a , \ocal^\dagger \ucal_1 \acal \ocal \acal^\dagger \ucal_1^\dagger \right] = 0 \,,
\label{gi.con}
\eea
where $ \ocal $ is evaluated at $y$.

For connected gauge groups one can always write
$ \ocal(y) = \exp( i \omega_a(y) \tau_a ) $; in this
case the first equation in (\ref{gi.con}) is
satisfied once (\ref{pcon}) is imposed. The second 
equation in (\ref{gi.con}) is also satisfied since
by (\ref{pcon}) $  \ucal_1 \acal $ commutes with all 
the $ \ocal $. So, the bosonic BC are gauge invariant as a consequence
of the symmetry of the Lagrangian under the twist operation (\ref{Pbc}) {\it and}
of the gauge symmetry of the fermionic BC (\ref{f_gi}).

Similar arguments for the other three types
of boundary conditions show that for any choice $Pi-Rj$
the theory retains its local symmetry provided (\ref{pcon}) 
and (\ref{rcon}) are valid and if the 
gauge transformations are restricted by the conditions
\beq
\ocal(y+L) = \acal \ocal(y) \acal^\dagger
\qquad {\rm and} \qquad
\ocal(-y) = \bcal \ocal(y) \bcal^\dagger\,.
\label{gauge.inv}
\eeq

For non-Abelian groups it is not too difficult (at least for infinitesimal transformations) 
to show that the converse, {\it i.e.} that
the gauge invariance of the bosonic BC (\ref{b_gi}) implies that the invariance of the fermionic ones (\ref{f_gi})
(which is equivalent to (\ref{gauge.inv}))
also holds, provided (\ref{pcon})  and (\ref{rcon}) are satisfied. In other words,
the bosonic BC are gauge invariant if and only if the fermionic BC are gauge invariant, 
provided the theory is symmetric under the twist operations  (\ref{pcon})  and (\ref{rcon}). For
Abelian groups a similar calculation leaves a phase
ambiguity. 

When the fields are expanded in Fourier series, the
conditions (\ref{pcon}) and (\ref{rcon}) often forbid the
presence of zero modes for some of $A_N^a$. 
The absence of certain gauge boson zero modes is directly related to constraints which
must be satisfied by the gauge functions $\omega_a(y)$ to obey (\ref{gauge.inv}).

For instance, as a prelude to
the discussion of various Abelian examples in section~\ref{onefer}, it is worth listing
here for a $\ui$ gauge theory the forbidden gauge-boson modes together with 
the restrictions on the allowed gauge transformation that follow from (\ref{gauge.inv}):
\bit
\item P1/P2: The gauge invariance of BC requires periodicity (P1) or anti-periodicity (P2); 
$\Lambda(y+L)=\pm\Lambda(y)$ ($\Lambda(y)$ is the
U(1) gauge function: $A_M\to A_M + \partial_M \Lambda$).  
Note that for P2 the anti-periodicity 
of $A_\mu(y)$ eliminates a massless photon for this choice of BC.

\item R1/R2: Here for the invariance of the BC  one needs $\Lambda(-y)=\pm \Lambda(y)$. 
In particular a massless gauge-boson mode is not allowed by the odd boundary condition R2..
\eit
Note that $y$-independent gauge transformations are not allowed for P2 or R2 
due, respectively, to the anti-periodicity or asymmetry of $\Lambda(y)$; 
in these cases the gauge symmetry of the zero-mode sector 
(i.e. KK modes of $y$-independent 5D fields)  is broken completely. This 
will be discussed in detail in section~\ref{lightfer.gi}.

Though the BC may reduce the gauge symmetry within the zero-mode sector,
{\it the whole theory remains 5D gauge invariant}. It is not difficult to show that, 
at least for infinitesimal gauge transformations, 
there exists a basis (in general different basis must be adopted for 
the periodicity and the orbifold conditions) such that (\ref{gauge.inv})
 reduces to $\omega_a(y+L)=\pm\omega_a(y)$ and
$\omega_a(-y)=\pm\omega_a(y)$ (signs are uncorrelated). Therefore, 
choosing appropriate values for $\omega(0)$ and $\omega(\pm L/2)$
it is always possible to find {\it all} non-zero, continuous and differentiable
$\omega_a(y)$ such that (\ref{gauge.inv}) is satisfied. S
The initial symmetry group remains unchanged since
{\it all} $\omega_a(y)$ are non-zero, though their functional
form is constrained by the above periodicity and reflection conditions..

For example, consider an \su2 theory with a single doublet
and (P1-R1) BC. Taking $ \Gamma= U_1 = i \sigma_3$, 
$ \tilde\Gamma = -i \tilde U_1 = \sigma_1$ and $ \Upsilon=
\tilde\Upsilon=0 $ (so $\acal,\bcal\neq\mati$) the conditions, ({\ref{gauge.inv}), on  the gauge
transformation functions $ \Omega = \exp( i \sigma^a \omega_a)$
imply
\beq
\begin{array}{ccc}
 \omega_1(y) = &- \omega_1(y+L) = &+ \omega_1(-y) \cr
 \omega_2(y) = &- \omega_2(y+L) = &- \omega_2(-y) \cr
 \omega_3(y) = &+ \omega_3(y+L) = &- \omega_3(-y)\,.
\end{array}
\label{examp}
\eeq
Therefore the theory (including the BC) will have a local \su2 symmetry
provided the $\omega_a(y)$ satisfy the above constraints.
If we  had chosen instead  
$ \Gamma= U_1 = \tilde\Gamma = \tilde U_1 = \mati, ~ \Upsilon=
\tilde\Upsilon=0 $ ($\acal=\bcal=\mati$) then the BC are gauge 
invariant provided $ \omega_a(y) = \omega_a(y+L) = 
\omega_a(-y) ~(a=1,2,3) $; since the $\omega_a$ are all non zero, this is again 
a local \su2 theory, but not with the same local group
as in the first case, in fact, the only common element is $ \Omega = \mati$.
This also illustrates another interesting fact, namely, that
non-trivial choices of $ \acal$ and $ \bcal $, 
{\it i.e.}  $ \acal\neq\mati$ and $ \bcal\neq\mati $,
do not reduce the 5D local symmetry group (as we have just argued the group remains the same), 
but it may simply change it as we have observed in the above example.

Let us briefly discuss the gauge symmetry of the zero-mode sector in the above example
(\ref{examp}). For $y-$independent transformations the periodicity condition P1
requires $ \omega_{1,2}=0$, while R1 requires $ \omega_3=0 $. In this case
the gauge group of the light sector is completely broken.

The above scheme of gauge symmetry breaking in the zero-mode sector by BC (the Scherk-Schwarz mechanism) could be also
viewed from the following perspective. The 5D gauge symmetry is associated with
a set of unconstrained gauge functions $\omega_a(y)$. Imposing 
BC restricts the set of allowed $\omega_a(y)$'s, for instance
requiring them to be anti-periodic and even. 
Therefore the symmetry is ``reduced'' by which we mean
that none of the generators is broken (none of the  $ \omega_a$
is required to vanish identically by the BC) and yet {\it the zero-mode
sector has only a subgroup of the original group}.
For instance, in a $\ui$ gauge model with R2,  $A_\mu(y)$
has no zero mode.

\subsection{General solutions for the allowed boundary conditions}
\label{gen_sol}

The conditions (\ref{pcon}) and (\ref{rcon}) significantly constrain
the form of $\acal$ and $ \bcal $. To derive the general structure 
of these matrices we decompose $ \Psi $ in terms of multiplets 
$ \psi_r$ being each in an \irrep\ $r$ of the gauge group. 
As a preliminary result we first
show that when $r$ is complex we can assume without loss
of generality that $ \psi $ contains no multiplet
transforming according to the complex-conjugate \irrep\
$ \bar r$.

To see this first note that given the
structure of the Lagrangian we can assume that the mass
matrix is diagonal, and we will denote by $m_r$ the eigenvalue
associated with $\psi_r$. Then if the theory does originally contains
a fermion multiplet $\psi_{\bar r}$ transforming according to the 
\irrep\ $\bar r$, the terms in $ \lcal $ where this field appears
are
\beq
\lcal\up{\bar r} = \bar\psi_{\bar r}
\left[
 i \gamma^N \left( \partial_N + i g_5 T_a\up{\bar r} A_N^a \right)
 - m_{\bar r} \right] \psi_{\bar r}\,,
\eeq
where the $ T_a\up{\bar r} $ generate the corresponding representation. 
It is then possible to define a field $ \psi_r' = ( \psi_{\bar r} )^c $ (that 
transforms according to the complex-conjugate \irrep\ $r$) in terms of which
\beq
\lcal\up{\bar r} = \bar\psi_r'
\left[
 i \gamma^N \left( \partial_N + i g_5 T_a\up r A_N^a \right)
 - m_r' \right] \psi_r' \,,
\eeq
where $ m_r' = - m_{\bar r} $ and  $ T_a\up r=
 - \left( T_a\up{\bar r}\right)^* $.
Since we can replace each $ \psi_{\bar r} $ by its
corresponding $ \psi_r'$, we can assume that $ \Psi $ 
contains no multiplets in the complex conjugate \irrep\
$ \bar r $. 

This way of eliminating conjugate representations does
not lead to any simplifications for real or pseudoreal
representations $ r_u$ since the corresponding generators satisfy
\beq
T_a\up{\bar r_u} = \left[ - T_a\up{r_u} \right]^* 
= S_u T_a\up{r_u} S_u^\dagger\,,
\eeq
for some unitary matrix $ S_u $ that is (anti)symmetric
for (pseudo)real representations.

The above arguments imply that we can choose fields such that
\beq
T_a = \diag\left( \cdots , \mati_{n_\ell} \otimes T_a\up{r_\ell} , \cdots, 
\mati_{n_u} \otimes T_a\up{r_u} , \cdots \right)\,,
\label{form.of.Ta}
\eeq
where we assume the theory contains $ n_\ell$ flavors
in the complex \irrep\ $ r_\ell$ and $n_u$ flavors 
in the (pseudo)real \irrep\ $ r_u $. In this expression
as in the rest of the paper a matrix of the form $\fcal \otimes \gcal$
is understood as having $\fcal$ ($\gcal$) act on the flavor (gauge) indices,
and $\mati_n $ denotes the $n \times n$ unit matrix.

Letting $ d_{\ell,u}$ be the dimension of $ r_{\ell,u}$ we define
\beq
F = \diag\left(\cdots , \mati_{n_\ell} \otimes \mati_{d_\ell} , \cdots, 
\mati_{n_u} \otimes \mati_{d_u} , \cdots ; 
\cdots , \mati_{n_\ell} \otimes \mati_{d_\ell} , \cdots, 
\mati_{n_u} \otimes S_u , \cdots  \right)\,,
\eeq
so that $ \tau_a = F \tau'_a F^\dagger$, where
\beq
\tau_a' = \diag\left( \cdots , \mati_{n_\ell} \otimes T_a\up{r_\ell} , \cdots, 
\mati_{n_u} \otimes T_a\up{r_u} , \cdots ;
          \cdots , \mati_{n_\ell} \otimes T_a\up{\bar r_\ell} , \cdots, 
\mati_{n_u} \otimes T_a\up{r_u} , \cdots \right)\,.
\eeq

Adopting the Schur's lemma and the grand orthogonality theorem
(used to eliminate the possibility that the $T_a\up r$
might be linearly dependent) the twist-invariance conditions 
$ [ \tau_a' , F^\dagger \ucal_i \acal F ] =0 $,
$ [ \tau_a' , F^\dagger \tilde\ucal_i \bcal F ] =0 $, 
imply that 
$ F^\dagger \ucal_i \acal F $ and
$  F^\dagger \tilde\ucal_i \bcal F $ have no entries connecting
two inequivalent representations, and that entries
connecting equivalent representations will be diagonal
in the gauge indices. Explicitly we obtain
\bea
F^\dagger \ucal_i \acal F &=& \left( \begin{array}{cccc}
X_\ell \otimes \mati_{d_\ell} & 0 & 0 & 0 \cr
0 & X_u \otimes \mati_{d_u}& 0 & Y_u' \otimes \mati_{d_u} \cr
0 & 0 & X'_\ell \otimes \mati_{d_\ell} & 0 \cr
0 & Y_u \otimes \mati_{d_u} & 0 & X_u' \otimes \mati_{d_u}
\end{array} \right) \cr
F^\dagger \tilde\ucal_i \bcal F &=& \left( \begin{array}{cccc}
\tilde X_\ell \otimes \mati_{d_\ell} & 0 & 0 & 0 \cr
0 & \tilde X_u \otimes \mati_{d_u} & 0 & \tilde Y_u' \otimes \mati_{d_u} \cr
0 & 0 & \tilde X'_\ell \otimes \mati_{d_\ell} & 0 \cr
0 & \tilde Y_u \otimes \mati_{d_u} & 0 & \tilde X_u' \otimes \mati_{d_u}
\end{array} \right) \,.
\eea

If $ U_i = \exp\{ i u^i_a T_a \}$, we denote by
$U_{i;\ell} = \exp\left\{i u_a^i T_a\up{r_\ell} \right\}, 
~i=1,2$ and similarly for $U_{i;u}$, $\tilde U_{i;\ell}$ 
and $\tilde U_{i;u}$. Then, using the unitarity of 
$ \ucal_i$ and $F$ we find
\beq
\begin{array}{rl}
P1: & \Gamma = \diag\left( \cdots , X_{1;\ell}\otimes U_{1;\ell}^\dagger ,
 \cdots ,  X_{1;u}\otimes U_{1;u}^\dagger, \cdots \right) \cr
   & \Upsilon = \diag\left( \cdots , 0 , 
           \cdots ,  Y_{1;u}\otimes U_{1;u}^T S_u, \cdots \right) \cr
P2: & \Gamma = \diag\left( \cdots , 0 , 
           \cdots ,  Y_{2;u}\otimes U_{2;u}^\dagger S_u, \cdots \right) \cr
   & \Upsilon = \diag\left( \cdots , X_{2;\ell}\otimes U_{2;\ell}^T ,
 \cdots ,  X_{2;u}\otimes U_{2;u}^T, \cdots \right) \cr
R1: &  \tilde\Gamma = -\diag\left( \cdots , 
   \tilde X_{1;\ell}\otimes \tilde U_{1;\ell}^\dagger ,
 \cdots ,  \tilde X_{1;u}\otimes \tilde U_{1;u}^\dagger, \cdots \right) \cr
   &  \tilde\Upsilon = \diag\left( \cdots , 0 , 
           \cdots ,  
    \tilde Y_{1;u}\otimes \tilde U_{1;u}^T S_u, \cdots \right) \cr
R2: & \tilde\Gamma = -\diag\left( \cdots , 0 , 
           \cdots ,  
   \tilde Y_{2;u}\otimes \tilde U_{2;u}^\dagger S_u, \cdots \right) \cr
   &  \tilde\Upsilon = \diag\left( \cdots , 
   \tilde X_{2;\ell}\otimes \tilde U_{2;\ell}^T ,
 \cdots ,  \tilde X_{2;u}\otimes \tilde U_{2;u}^T, \cdots \right) \,.
\end{array}
\label{gi.and.AB}
\eeq
The specific form of $ \acal $ and $ \bcal $ in (\ref{adef}) and (\ref{bdef}) allows
$ X_u',~ Y_u',~ \tilde X_u', \tilde Y_u' $ to be written in terms of
$ X_u,~ Y_u,~ \tilde X_u, \tilde Y_u $, but these relations will not
be displayed as they are not needed.

The unitarity of $ \acal $ and $ \bcal $ implies:
\beq
\begin{array}{ll}
X_{i;\ell}^\dagger X_{i;\ell} = \mati_{n_\ell} 
& \tilde X_{i;\ell}^\dagger \tilde X_{i;\ell} = \mati_{n_\ell} \cr
X_{i;u}^\dagger X_{i;u} + Y_{i;u}^\dagger Y_{i;u} = \mati_{n_u} 
& \tilde X_{i;u}^\dagger \tilde X_{i;u} + 
       \tilde Y_{i;u}^\dagger \tilde Y_{i;u} = \mati_{n_u} \cr
X_{i;u}^T Y_{i;u} = \pm Y_{i;u}^T X_{i;u} 
& \tilde X_{i;u}^T \tilde Y_{i;u} = \pm \tilde Y_{i;u}^T \tilde X_{i;u} \,,
\end{array}
\label{a.b.unitarity}
\eeq
where the upper(lower) signs corresponds to (pseudo)real
\irrep s.

The consistency condition $ \bcal = \bcal^\dagger $ requires,
for complex representations,
\beq
\begin{array}{rlll}
R1: 
        & \tilde X_{1;\ell} = \tilde c_\ell \tilde X_{1;\ell}^\dagger
        & \tilde U^2_{1;\ell}  = \tilde c_\ell \mati_{d_\ell} \quad \left| \tilde c_\ell \right|^2 =1 \cr
R2: 
        & \tilde X_{2;\ell} = \tilde c_\ell \tilde X_{2;\ell}^T
        & \tilde U_{2;\ell} \tilde U^*_{2;\ell} = \tilde c_\ell \mati_{d_\ell} \quad \tilde c_\ell^2 =1 \,,\cr
\end{array}
\label{comp.conds}
\eeq
while for real or pseudoreal representations we find
\beq
\begin{array}{rlll}
R1: 
        & \tilde X_{1;u} = \tilde c_u \tilde X_{1;u}^\dagger
        & \tilde Y_{1;u} = \pm \tilde c_u \tilde Y_{1;u}^T 
        & \tilde U^2_{1;u}  = \tilde c_u \mati_{d_u}  \quad \tilde c_u^2 =1 \cr
R2: 
        & \tilde Y_{2;u} = \tilde c_u \tilde Y_{2;u}^\dagger
        & \tilde X_{2;u} = \tilde c_u \tilde X_{2;u}^T 
        & \tilde U_{2;u} \tilde U^*_{2;u} = \tilde c_u \mati_{d_u}  \quad \tilde c_u^2 =1 \,,\cr
\end{array}
\label{real.conds}
\eeq
where the (lower) upper sign refers to a (pseudo)real representation.
For R2 we used the fact that we can assume 
$ S_u^2 = \mati $~\footnote{For a non-complex representation (dropping the
$u$ subscript) and taking a basis where $C_i$ are the Cartan generators
and $ E_\alpbf$ the root generators, the conjugate representation is
generated by $ C_i' = S C_i S^\dagger = - C_i,~E'_\alpbf = 
S E_\alpbf S^\dagger = - E_{-\alpbf}$ from which it follows that
$S^2$ commutes with all the generators and so $ S^2 = \sigma \mati $
for some complex number $\sigma$, $|\sigma|=1$. Redefining $ S \to S/\sqrt{\sigma}$
shows we can take $ S^2 = \mati $.}. 
Notice that the above constraints on the matrices $U_{i;r}$ are sufficient to 
obey the gauge-boson-consistency  constraints (\ref{bb.b}).

The constraints required by $  \acal\bcal = (\acal\bcal)^\dagger $ 
can be obtained in a similar way. Let us define the matrices
\beq
\begin{array}{|l|ll||l|}
\hline \multicolumn{1}{|c|}{\hbox{BC}} & 
       \multicolumn{2}{c||}{\hbox{Non-complex}} &
       \multicolumn{1}{c|}{\hbox{Complex}}\\ \hline
P1-R1\quad & K_{11}\up u =-X_{1;u}\tilde X_{1;u} \pm Y_{1;u}^*\tilde Y_{1;u}    &
              \quad L_{11}\up u = X_{1;u}^* \tilde Y_{1;u} + Y_{1;u} \tilde X_{1;u}     &
                K_{11}\up \ell = - X_{1;\ell}\tilde X_{1;\ell}                  \\
P1-R2\quad & K_{12}\up u = -X_{1;u}\tilde Y_{2;u} \pm Y_{1;u}^*\tilde X_{2;u}   &
              \quad L_{12}\up u = X_{1;u}^* \tilde X_{2;u} + Y_{1;u} \tilde Y_{2;u}     &
                L_{12}\up \ell = X_{1;\ell}^* \tilde X_{2;\ell}                 \\
P2-R1\quad & K_{21}\up u = -Y_{2;u}\tilde X_{1;u} + X_{2;u}^*\tilde Y_{1;u}     &
              \quad L_{21}\up u = \pm Y_{2;u}^* \tilde Y_{1;u} + X_{2;u} \tilde X_{1;u} &
                L_{21}\up \ell = X_{2;\ell} \tilde X_{1;\ell}           \\
P2-R2\quad & K_{22}\up u = -Y_{2;u}\tilde Y_{2;u} + X_{2;u}^*\tilde X_{2;u}     &
              \quad L_{22}\up u = \pm Y_{2;u}^* \tilde X_{2;u} + X_{2;u} \tilde Y_{2;u} &       
                K_{22}\up \ell = X_{2;\ell}^*\tilde X_{2;\ell}                  \\ \hline
\end{array} \,,
\label{k.l.defs}
\eeq
where the upper (lower) signs refer to (pseudo)real representations.
In order to fulfill the condition $  \acal\bcal = (\acal\bcal)^\dagger $ these matrices should satisfy
\beq
K_{ij}\up r = \lambda_{ij}\up r K_{ij}\up r{}^\dagger \qquad 
L_{ij}\up r = s_{ij}\up r L_{ij}\up r{}^T ; \quad \left(s_{ij}\up r\right)^2=1
\label{her.sym}
\eeq
(no sum over $i,j=1,2$). For complex representations 
$ \left|\lambda_{ij}\up\ell \right| = 1 $
while for  non-complex representations
\beq
\begin{array}{ll}
\lambda_{11}\up u = \pm s_{11}\up u & \lambda_{12}\up u = s_{12}\up u \cr
\lambda_{22}\up u = \pm s_{22}\up u & \lambda_{21}\up u = s_{21}\up u \,,\cr
\end{array}
\label{def.of.s.lam}
\eeq
where the upper (lower) sign refers to a (pseudo)real representation.
The corresponding restrictions on the matrices $ U_{i;r} $ are
\beq
\begin{array}{|l|l|}\hline
P1-R1\quad & \tilde U_{1;r} U_{1;r} = \lambda_{11}\up r U_{1;r}^\dagger \tilde U_{1;r}^\dagger \cr
P1-R2\quad & \tilde U_{2;r} U_{1;r} = s_{12}\up r U_{1;r}^T \tilde U_{2;r}^T \cr
P2-R1\quad & \tilde U_{1;r}^* U_{2;r} = s_{21}\up r U_{2;r}^T \tilde U_{1;r}^\dagger \cr
P2-R2\quad & \tilde U_{2;u}^* U_{2;r} = \lambda_{22}\up r U_{2;r}^\dagger \tilde U_{2;r}^T \\ \hline
\end{array}\,.
\label{babaU}
\eeq
It is worth noting that the consistency  conditions (\ref{baba.b})
also lead to constraints of the form (\ref{babaU}) but with
$ \lambda_{ij}\up r,~s_{ij}\up r $ arbitrary complex numbers; 
the additional restrictions (\ref{her.sym}), (\ref{def.of.s.lam}) follow
exclusively from the Hermiticity of $\acal\bcal$.

\bigskip

The expressions 
(\ref{gi.and.AB})
(\ref{a.b.unitarity})
(\ref{comp.conds})
(\ref{real.conds}) and 
(\ref{babaU}) together with $ \left| \lambda_{ij}\up\ell \right|=1$ and (\ref{def.of.s.lam})
give the most general form for the matrices $\acal$, $\bcal$, $\ucal_i$
and $ \tilde\ucal_j$. In particular
\bit
\item The matrices $ \Gamma, ~\Upsilon, ~\tilde\Gamma$ and 
$\tilde\Upsilon$ do not mix $ \psi_r $ and $\psi_s$ unless
$r$ is equivalent to $s$ or $\bar s$.
\item For the BC P1 and R1 (P2 and R2), in the subspace spanned by all multiplets
in the same {\em complex} \irrep\ $r$, the matrices
$ \Upsilon$ and $ \tilde\Upsilon$ ($\Gamma$ and $\tilde\Gamma$) vanish. In contrast,
$ \Gamma$ and $ \tilde\Gamma$ ($ \Upsilon$ and $ \tilde\Upsilon$)
are direct products of unitary rotation in flavor indices and global gauge transformation in gauge
indices.
\item In the subspace spanned by all multiplets
carrying the same {\em (pseudo) real} \irrep\ $r$ in general
$ \Upsilon$ ($\tilde \Upsilon$) and $ \Gamma $ ($\tilde \Gamma $) are non-zero.
\eit

One of the virtues of including the generalized twist operations
is that they allow {\em all} mixing consistent with gauge invariance;
a more restricted standard set ($\Upsilon=\tilde\Upsilon=0$) of BC would not, for example, allow a
mixing between $ \psi_r $ and $ \psi_{r'}^c$ even though they
might transform in the same way under the local symmetry group.
The price we pay for this generalization is the breaking by the BC
of global fermion  number (for non-complex representations), and
possibly other global symmetries.

There is a comment here in order. The twist invariance conditions,
(\ref{pcon}) and (\ref{rcon}), guarantee that the Lagrangian is symmetric
under the twist operations defined by (\ref{Pbc}) and (\ref{Rbc}).
In addition the fermionic and bosonic twist operations must satisfy
the consistency conditions, (\ref{baba.f}-\ref{bb.f}) and (\ref{baba.b}-\ref{bb.b}), 
respectively. It is interesting to observe that our general 
solutions (\ref{real.conds}) and (\ref{babaU}) show that in fact the fermionic consistency
conditions (\ref{baba.f}-\ref{bb.f}) {\it imply} that
the bosonic ones (\ref{baba.b}-\ref{bb.b}) are satisfied. This remarkable
fact has been confirmed in all the examples considered in
sections \ref{sect:abelian.case} and \ref{sect:nonabelian.case}; 
the solutions for $\ucal_i$ obtained
by imposing the fermionic consistency condition automatically satisfy
the bosonic ones.

\subsection{\Vev s}

One property exhibited by many 5D systems is the
possibility that $A_4$ may acquire a \vev, which can lead to a variety
of interesting consequences such as spontaneous breaking of CP \cite{Grzadkowski:2004jv}.
In order to determine the constraints imposed on such a \vev\
by the various boundary conditions described above we define
\beq
\aBB \equiv \left(\begin{array}{cc}
                \vevof{A_4} & 0 \cr
                 0 & -\vevof{A_4}^*
             \end{array} \right)\,,
\eeq
which is preserved by $Pi-Rj$ provided
\beq
[ \aBB, \ucal_i ] = 0 \qquad \{ \aBB , \tilde\ucal_j \} =0 \,.
\label{bc.vev}
\eeq
For a non-Abelian group there are always non-trivial solutions to
these equations. For an Abelian groups, however, only the case $P1-R2$
allows a non-zero \vev. Note also that even if a non-zero
\vev\ is allowed this does not imply that such a $ \vevof{A_4}$ 
will correspond to absolute minima of the effective potential;
this can be only decided by explicitly calculating the effective
potential and will depend on the fermion content of the theory.

\subsection{C, P and CP}

In 5D the parity transformation acting on the space-time points is defined as follows:  
$x^{0,4}\to x^{0,4}$ and $x^i\to -x^i$, for $i=1,2,3$. Therefore for the parity
acting on fermionic fields and for the charge conjugation we obtain:
\beq
\chi \; {\stackrel P \longrightarrow} \; \gamma_0 \gamma_4 
\chi\,, \quad
\chi\; {\stackrel C \longrightarrow}\; \chi^c = 
\left(
\baa{cc}0&-1\\1&0\eaa
\right)
\chi\,.
\label{CandP}
\eeq
Then under CP we obtain
\beq
\chi \; {\stackrel {CP} \longrightarrow} \; \gamma_0 \gamma_4 \dcal \chi \equiv
\gamma_0 \gamma_4 
\left(
\baa{cc}0&-1\\1&0\eaa
\right)
\chi
\label{cp}
\eeq
while the gauge fields transform as
\beq
A_i  {\stackrel {CP} \longrightarrow}  + A_i^T \,,
\qquad
A_{0,4}  {\stackrel {CP} \longrightarrow}  -  A_{0,4}^T \,,
\eeq
where $i=1,2,3$.

One can generalize these definitions by noting that 
the kinetic term in $ \lcal $ is invariant under
unitary flavor mixing~\cite{Lee:1966ik} among fields belonging to the same 
representation, so in (\ref{cp}) we can replace
\beq
\dcal=
\left(
\baa{cc}0&-\Theta^*\\\Theta&0\eaa
\right) \quad {\rm with} \quad \Theta^\dagger\Theta=\mati\,.
\eeq
In that general case we find
\beq
A_i  {\stackrel {CP} \longrightarrow}  + \Theta A_i^T\Theta^\dagger \,,
\qquad
A_{0,4}  {\stackrel {CP} \longrightarrow}  -  \Theta A_{0,4}^T\Theta^\dagger \,,
\eeq
where $i=1,2,3$.

The condition for the invariance of the mass term under 
CP is $\{\mcal,\dcal\}=0$, or, equivalently,
$M\Theta^*=\Theta^*M^*$. Since $M$ is Hermitian, we can always adopt a basis
where it is real and diagonal, in which case this condition 
reduces to $ [ M,\Theta] =0 $. It
follows that in the absence of CP violation (CPV) we can find a basis where both $M$
and $\Theta$ are diagonal; one can then choose the field
phases such that the matrix $ \dcal $ is given by the 
simple expression used in (\ref{cp}).
If, on the other hand $ \Theta $ is such that (in a basis
where $M$ is real and diagonal) $ [ M , \Theta ] \not=0 $,
then the mass term will explicitly break CP.

The boundary conditions will preserve CP invariance only if
\beq
\begin{array}{ll}
[\acal^* , \dcal ]  =0 & U_i = U_i^* ~(i=1,2)\cr
\{\bcal^* , \dcal \} =0 & \tilde U_i = \tilde U_i^* ~(i=1,2)\cr
\label{cpcon}
\end{array}
\eeq
or, equivalently,
\bea
\Theta^\dagger \Gamma \Theta = \Gamma^*\,,  \qquad 
\Theta^\dagger \tilde\Gamma \Theta = \tilde\Gamma^* \,,\qquad 
\Theta^T \Upsilon \Theta = \Upsilon^*\,, \qquad 
\Theta^T \tilde\Upsilon \Theta = \tilde\Upsilon^*\,,
\label{no.cpv}
\eea
which, for $ \Theta = \mati $, merely requires $ \acal $ and $ \bcal $ to be
real.
If any one of these conditions is violated the boundary conditions 
will break CP explicitly.

This theory also contains a third source of CPV: the \vev of $A_4$.
If $ \vevof{A_4} \not = 0$ then we can take this matrix as proportional
to a Cartan generator\footnote{That is, there is a group rotation that
takes this matrix into a Cartan generator times a real number}
which is a symmetric matrix; then 
\beq
A_4 {\stackrel {CP} \longrightarrow} - A_4\,,
\eeq
so that such a \vev\ violates CP spontaneously.

Summarizing: the theory described by (\ref{lag}) will respect CP only
if $ [ M , \Theta] = \vevof{A_4} = 0 $ and if the conditions (\ref{cpcon})
are obeyed.

\section{Abelian examples}
\label{sect:abelian.case}

In this section we will illustrate the consequences of the above
generalized boundary conditions for the case of an Abelian group.
We first study the case of a single fermion and then that of two
fermions that exhibits some new features.

\subsection{One fermion}
\label{onefer}

For a single fermion of mass $m$ and charge $q$ (in units of $g_5$) we have 
$ \mcal = m \sigma_3 $ and $ \tau = q \sigma_3 $ (there is a single
group generator so we drop the subindex $a$). Imposing the previous
constraints on $ \acal$ and $ \bcal $ and using the freedom to choose
the global phase of the fields to eliminate some of the phases, we find
the following expressions,
\beq
\begin{array}{l|c|c|c|c}
      & P1-R1       & P1-R2                   & P2-R1         & P2-R2       \cr
\hline
\acal & s_a{\bf1}   & u {\bf1} + i v \sigma_3 &  - i \sigma_2 &-i\sigma_2    \cr
\bcal &-s_b\sigma_3 &  \sigma_1               &-s_b\sigma_3   & s_b \sigma_1 
\end{array}
\label{one.f.p.r}
\eeq
where $ u^2+v^2 =1 $ and $s_{a,b}^2 = 1 $ (the signs $s_a$ and $s_b$ are 
uncorrelated). The bare-mass term in $ \lcal $ and a possible non-zero
\vev\ $\vevof{A_4} $ are allowed only by the combination $P1-R2$; in this
case the mass term will conserve CP.

Whenever BC involving P2 are chosen the fermion field obeys the periodicity 
condition $ \Psi(y+L) = \Psi^c(y)$, which leads to $ \Psi(y+2L) = - \Psi(y) $. Then
\bea
P2: &&
\Psi(y) = \inv{\sqrt{2L}} \sum_{n=-\infty}^\infty e^{i(n+1/2)\pi y/L} \psi_n \cr
&& \psi_n = e^{-i (n+1/2) \pi} \psi^c_{-n-1}\,.
\eea
If we also impose R1 this expression is further constrained by 
$ \Psi(-y) = s_b \gamma_5 \Psi(y)$, as a result we find
\beq
\psi_n = \left( \begin{array}{cc}
                    s_b (-1)^n \sigma_2 \varphi_n^* \cr \varphi_n
                \end{array} \right) , \quad
\varphi_{-n-1} = (-1)^n \sigma_2 \varphi_n^*
\qquad (P2-R1)\,,
\eeq
where $ \varphi_n$ is a 2-component spinor. For these boundary conditions
($P2-R1$) a bare-mass term for the fermion is not allowed in the Lagrangian,
nonetheless the $ \varphi$ receive a Majorana mass of order $1/L$ from the kinetic
terms:
\beq
\int_0^L dy \bar\Psi \gamma^4 i \partial_4 \Psi = \sum_{n=0}^\infty
\frac{ i s_b (-1)^n \pi(2n+1)}{2L} \varphi_n^T \sigma_2 \varphi_n + \hbox{H.c.}
\qquad (P2-R1)\,.
\eeq
Similar results are obtained for $ P2-R2 $:
\bea
\psi_n &=& \left( \begin{array}{cc}
                    -i s_b \sigma_2 \varphi_n^* \cr \varphi_n
                \end{array} \right) , \quad
\varphi_{-n-1} = (-1)^n \sigma_2 \varphi_n^* \cr
\int_0^L dy \bar\Psi \gamma^4 i \partial_4 \Psi &=& -\sum_{n=0}^\infty
\frac{ s_b  \pi(2n+1)}{2L} \varphi_n^T \sigma_2 \varphi_n + \hbox{H.c.}
\qquad (P2-R2)\,.
\eea

If, on the other hand, we impose the periodicity condition P1 then
$\Psi(y+L ) = e^{i \alpha} \Psi(y)$ 
(corresponding to $\cos\alpha \equiv u$, $\sin\alpha \equiv v$ in Eq. \ref{one.f.p.r})
and we can write
\bea
P1:&& 
\Psi(y) = \inv{\sqrt{L}} \sum_{n=-\infty}^\infty \psi_n e^{i (2\pi n + \alpha)y/L}\,.
\label{p1r2_exp}
\eea
If we also impose R2 then $ \psi_n $ must obey 
$ \psi_n = \gamma_5 \psi_n^c$ and this leads to 
\beq
\psi_n = \left( \begin{array}{cc}
                    -i \sigma_2 \varphi_n^* \cr \varphi_n
                \end{array} \right) \qquad (P1-R2)\,.
\label{p1r2.bc}
\eeq
In this case the bare mass fermion term in the Lagrangian is allowed
since under the orbifold twist transformation, $\psi \to \gamma_5\psi^c$, the
5D fermion mass term is invariant:
\beq
\bar\psi\psi \to - \bar\psi^c\psi^c=\bar\psi\psi\,.
\label{fmass}
\eeq
In addition, the kinetic term also generates a Majorana mass term:
\bea
&& \hspace{-.5in}
\int_0^L dy \bar\Psi \left( i \gamma^4 \partial_4 - M \right) \Psi = \cr
&& \hspace{-.2in} = \sum_{n=-\infty}^\infty\left[
2 M \varphi_n^\dagger \varphi_n
- \frac{ ( 2 \pi n + \alpha)}L \left( 
\varphi_n^T \sigma_2 \varphi_n + \hbox{H.c.} \right) \right]
\qquad (P1-R2)\,.
\label{p1-r21fer}
\eea
When present, the \vev\ $ \vevof{A_4}$ will generate an
additional contribution to the mass. This important case is described
in more detail in the Appendix.

Finally, for the remaining $P1-R1$ case we find 
\bea
\Psi(y) &=& s_a \inv{\sqrt{L}} \sum_{n=-\infty}^\infty \psi_n e^{2\pi i n y/L} ;
\qquad \psi_{-n} = s_b \gamma_5 \psi_{n-\nu} \cr
\int_0^L dy \bar\Psi \gamma^4 i \partial_4 \Psi &=& 
\sum_{n=1-\nu}^\infty i \frac{(2 n+\nu)\pi}L \bar\psi_n \gamma_5 \psi_n
\qquad (P1-R1)\,,
\eea
where $\nu = (1-s_a)/2 $.  

It is worth pointing out that massless fermions are present only
when (P1-R1) BC are imposed with $ s_a=+1 $ or for (P1-R2) if $M=0$ and $\alpha=0$.
Note also that the (P1-R1) case is the only one where KK fermions are not restricted to be Majorana
fermions. This is related to the fact that these BC are invariant with respect to {\it global} U(1)
rescaling of the 5D fermion field $\Psi(y)$;  only for this choice fermion number remains conserved.
In all cases containing P2 and/or R2
\bit
\item P2: $\Psi(y+L) \propto \Psi^c(y)$, 
\item R2: $\Psi(-y) \propto \Psi^c(y)$,
\eit
so any global U(1) symmetry is broken by the fermionic BC. 

A common consequence of fermion number breaking 
is the generation of Majorana mass terms and this indeed occurs above; 
the KK fermionic modes are then 4D Majorana fermions. 
Therefore the generalized BC discussed in this paper provide a natural
method of constructing 4D Majorana fermions with masses of order $1/L$. 
This can be useful when building a realistic
models for neutrino interactions, especially if a seesaw mechanism is
also implemented.

\subsection{Two Abelian fermions}
\label{twofer}

The case for two Abelian fermions can be studied along similar
lines. In this case we have $M = \hbox{diag}(m_1,m_2)$, 
$T= \hbox{diag}(q_1,q_2)$~\footnote{The U(1) gauge 
symmetry implies that $[M,T]=0$, so therefore both 
$M$ and $T$ can be chosen diagonal.} and
\beq
\Psi = \left( \begin{array}{c} \psi_1 \cr \psi_2 \end{array} \right)\,.
\eeq
We assume that masses and charges are not zero.
The richest structure is allowed
by the $P1-R2$ BC which we consider first. 

\subsubsection{P1-R2 boundary conditions}

This case has a special interest because it is the only
one (for the two-Abelian fermion model) 
that allows a non-zero \vev\ for $ A_4 $. The 
BC for the gauge fields are
\beq
\baa{ll}
A_N(y+L)=A_N(y)&\\
A_\mu(-y)=-A_\mu(y)&A_4(-y)=A_4(y)\,.
\label{bosonbc}
\eaa
\eeq
Concerning the fermions, the conditions
$ [\acal,\tau]=\{\bcal,\tau\}=0$ and
$ [\acal, \mcal] = \{\bcal,\mcal \} = 0 $ suggest
that the cases where $ |m_1/m_2| = |q_1/q_2| =  1 $
should be treated separately. This, however is not the 
case.

Suppose, for example, that $ m_1/m_2 = q_1/q_2 = + 1 $,
then the constraints on $ \acal $ and $ \bcal $
imply $ \tilde\Gamma = \Upsilon = 0 $. In addition,
the Lagrangian has a $U(2)$ flavor symmetry that
allows us to choose $ \Gamma$ diagonal and $ \tilde\Upsilon = \mati$.
The boundary conditions then reduce to 
\beq
\psi_i(y+L) = e^{i \alpha_i} \psi_i(y), \quad
\psi_i(-y) = \gamma_5 \psi_i^c(y) 
\qquad (i=1,2).
\label{2f.bc}
\eeq
 so that each flavor 
has an expansion of the form (\ref{p1r2_exp})--(\ref{p1r2.bc}).
If $ m_1/m_2 = q_1/q_2 = - 1 $, then, following
the discussion of section \ref{gen_sol} it is convenient to 
introduce $ \psi_2 = (\psi_2')^c $ so that the new field
has mass and charge $ m_2' = - m_2 = + m_1$,
$ q_2' = - q_2 = + q_1 $; in terms of $ \psi_1,~\psi_2'$
the theory is identical to the one just considered.
If the masses and charges do not satisfy 
$ |m_1/m_2| = |q_1/q_2| =  1 $ then gauge invariance
requires that the boundary conditions be again given by
(\ref{2f.bc}).

We conclude that with appropriate choice of fields the two
fermions decouple from each other when the boundary conditions (P1-R2) are
imposed. In this case the considerations of the previous
section determine the physics of the model. In particular,
for this choice of BC, CP
is violated either explicitly non-zero $\alpha_i$ or spontaneously
by one-loop \vev\ of $A_4$, see the Appendix.

\subsubsection{P1-R1 boundary conditions}

In this case the constraints are satisfied only when $ m_1 = m_2$
and $ q_1 = - q_2 $ and provided $ \tilde\Gamma = \Upsilon =0 $,
$ \tilde\Upsilon = \lambda \sigma_1 $ with $ | \lambda| = 1 $
and $ \Gamma = \hbox{diag}( e^{i \alpha} , e^{- i \alpha })$ 
The freedom to redefine the  global phase of the fields can 
then be used to set $\lambda = 1 $, then we have
\beq
\begin{array}{ll}
\psi_1(y+L) = e^{+i\alpha} \psi_1(y) & \psi_1(-y) = \gamma_5 \psi_2^c(y) \cr
\psi_2(y+L) = e^{-i\alpha} \psi_2(y) & \psi_2(-y) = \gamma_5 \psi_1^c(y)\,.
\end{array}
\eeq
This can be used to eliminate $ \psi_2 $. The action then becomes
twice the action for $ \psi_1 $ alone, with $ \psi_1$ obeying
the above periodicity condition; this case also reduces to a single-fermion model.

\subsubsection{P2-R1 boundary conditions}

In this case the constraints are again satisfied only when $ m_1 = m_2$
and $ q_1 = - q_2 $ and provided $ \tilde\Gamma = \Upsilon =0 $,
$ \tilde\Upsilon = \lambda \sigma_1 $ with $ | \lambda| = 1 $
and 
\beq
\Gamma = \left(\begin{array}{cc} 0 & e^{i\beta} \cr e^{i\beta'} & 0 \end{array} \right)\,.
\eeq
The freedom to redefine the  global phase of the fields can 
then be used to set $\lambda =1,~ \beta' = 0 $, then we have
\beq
\begin{array}{ll}
\psi_1(y+L) = e^{+i\beta} \psi_2(y) & \psi_1(-y) = \gamma_5 \psi_2^c(y) \cr
\psi_2(y+L) = \psi_1(y)             & \psi_2(-y) = \gamma_5 \psi_1^c(y)\,.
\end{array}
\eeq
Using this to eliminate $ \psi_2 $, the action again becomes
twice the action for $ \psi_1 $ alone, where $ \psi_1$ obeys
$ \psi_1(y) = \gamma_5 \psi_1^c(-y-L) $ and $ \psi_1(y+2L) = 
e^{i\beta} \psi_1(y)$. Solving these yields
\beq
\psi_1(y) = \inv{\sqrt{2L} } \sum e^{ i \tilde\omega_n y}
\left( \begin{array}{c} e^{i [\beta  + (2n-1)\pi]/2} \sigma_2 \varphi_n^* 
\cr \varphi_n \end{array} \right); \quad \tilde\omega_n = \frac{2\pi n + \beta}{2L}\,.
\eeq

\subsubsection{P2-R2 boundary conditions}

These constraints require $m_1/m_2 = - q_1/q_2 = \pm 1 $. When
$ m_1 = m_2 $ we again find $ \tilde\Gamma = \Upsilon =0 $, and
using the the freedom to redefine the global phases allows us
to choose $ \Gamma = e^{i \beta} \sigma_1,~\tilde\Upsilon = {\bf1}$.
The boundary conditions then become
\beq
\begin{array}{ll}
\psi_1(y+L) = e^{+i\beta} \psi_2(y) & \psi_1(-y) = \gamma_5 \psi_1^c(y) \cr
\psi_2(y+L) = e^{+i\beta} \psi_1(y) & \psi_2(-y) = \gamma_5 \psi_2^c(y)\,.
\end{array}
\eeq
Again $ \psi_2 $ can be eliminated and the action then becomes
twice the action for $ \psi_1 $ alone; here the constraints on
$ \psi_1$ give the expansion
\beq
\psi_1(y) = \inv{\sqrt{2L} } \sum e^{ i \tilde\omega_n y}
\left( \begin{array}{c} (-1)^{n+1} i \sigma_2 \varphi_n^* 
\cr \varphi_n \end{array} \right); \quad \tilde\omega_n = \frac{\pi n + \beta}L\,.
\eeq

When $ m_1 = - m_2 $ similar arguments lead to 
$ \Gamma = \tilde\Gamma = 0 $, $ \Upsilon = e^{i \beta} \sigma_1,
~\tilde\Upsilon = {\bf1}$, and
\beq
\psi_1(y) = \inv{\sqrt{2L} } \sum e^{ i \tilde\omega_n y}
\left( \begin{array}{c} -i \sigma_2 \varphi_n^* 
\cr \varphi_n \end{array} \right); \quad \tilde\omega_n = \frac{\pi (n+1/2) + \beta}L\,.
\eeq
As in the previous cases we can use the BC to eliminate $ \psi_2 $, now
in terms of a {\em translated} $ \psi_1$: $ \psi_2 ( y ) = e^{i\beta} \psi_1^c(y+L)$.
The action then reduces to that for $ \psi_1$ alone, but with the radius
of the compact dimension equal to $2L$.

\section{Simple non-Abelian cases}
\label{sect:nonabelian.case}

\subsection{\su2 models}
\label{su2}

We consider a model with \su2 as the gauge group and where all fermions
transform according to the fundamental representation. This is a
pseudoreal representation generated by the Pauli matrices $ \sigma_I $; $ \sigma_2 $
plays the role of the matrix $S_u$ of section \ref{gen_sol}.

\subsubsection{One doublet}

When the theory contains a single \su2 doublet, $ X,~Y$, etc. of section \ref{gen_sol} are just 
numbers, then since the representation is 
pseudoreal, (\ref{gi.and.AB}) and (\ref{a.b.unitarity}) imply $ 
X_i Y_i = \tilde X_i \tilde Y_i =0 $ (we drop the representation index $u$),
which implies that for this case either $ \Gamma $ or
$\Upsilon $ vanish (similar conclusions can be drawn for
$ \tilde\Gamma $ and $ \tilde \Upsilon $). This leads to the following
possibilities
\beq
\begin{array}{|c||c|c|}
\hline
   & \Gamma                     & \Upsilon           \cr
\hline
P1_a & X_1 U_1^\dagger          & 0                  \cr
P1_b & 0                        & Y_1 U_1^T \sigma_2 \cr
\hline
P2_a & Y_2 U_2^\dagger \sigma_2 & 0                  \cr
P2_b & 0                        & X_2 U_2^T          \cr
\hline
\end{array}
\qquad
\begin{array}{|c||c|c|c|}
\hline
     & \tilde\Gamma & \tilde\Upsilon & \tilde U_j\cr
\hline
R1_a & -\tilde X_1 \tilde U_1^\dagger & 0 & \tilde X_1^2 \tilde U_1^\dagger \cr
R1_b & 0 & \tilde Y_1 \tilde U_1^T \sigma_2  & - \tilde U_1^\dagger \cr
\hline
R2_a & -\tilde Y_2 \tilde U_2^\dagger \sigma_2 & 0 & \tilde Y_2^2 \tilde U_2^T  \cr
R2_b & 0 & \tilde X_2 \tilde U_2^T & \tilde U_2^T\cr
\hline
\end{array}
\eeq
where $ |X_i| = |Y_j| = |\tilde X_k| = | \tilde Y_l|=1 $
and the last column in the $Rj$ table
gives the constraints on the matrices $ \tilde U_i$ imposed by the
consistency condition $ \bcal = \bcal^\dagger $.

The remaining consistency condition,  $ \acal \bcal = (\acal\bcal)^\dagger $ 
requires  $ \Gamma\tilde\Gamma + \Upsilon^*\tilde\Upsilon $ to be Hermitian
and  $\Gamma^*\tilde\Upsilon - \Upsilon\tilde\Gamma$ symmetric and leads to the
following constraints
\bea
Pi_r-Rj_s:&& (\tilde U_i^* U_i)^2 = \left\{ \begin{array}{ll}
(\tilde\lambda\lambda)^2 \mati_2& r=s \cr
- \mati_2& r\not=s
\end{array}  \right. \cr
P1_r-R2_s :&& \left\{
\begin{array}{ll}
(\sigma_2 \tilde U_2 U_1)^2 = (\lambda \tilde\lambda)^2 & r=s \cr
\tilde U_2 U_1 = (\tilde U_2^* U_1)^T                  & r\not=s
\end{array} \right. \cr
P2_r-R1_s :&& \left\{
\begin{array}{ll}
(\tilde U_1 \sigma_2 U_2)^2 = (\lambda^* \tilde\lambda)^2 & r=s \cr
\tilde U_1 U_2 = (\tilde U_1^* U_1)^T                & r\not=s
\end{array} \right.
\eea
where
\beq
\begin{array}{|c|c|c|c|c|}
\hline
         & P1_a & P1_b  & P2_a & P2_b  \cr \hline
\lambda: & X_1  & Y_1^* & Y_2  & X_2^* \cr
\hline \end{array}
\qquad
\begin{array}{|c|c|c|c|c|}
\hline
                & R1_a        & R1_b         & R2_a        & R2_b \cr \hline
\tilde \lambda: & -\tilde X_1 & \tilde Y_1^* & -\tilde Y_2 & \tilde X_2^* \cr
\hline \end{array}\,.
\eeq

\subsubsection{Two \su2 doublets}

We have shown in (\ref{a.b.unitarity}) that for pseudoreal representations
$\Gamma$ ($\tilde\Gamma$) and $\Upsilon$ ($\tilde\Upsilon$)
can be simultaneously non-zero only if at least two \su2 pseudo-real multiplets 
are present; this section illustrates such a scenario.
For simplicity we will restrict ourselves to the case of only two doublets.

Using then the
freedom to make unitarity rotations of the doublets (which might render
a non-diagonal mass matrix) it is straightforward to show that the
rest of the conditions (\ref{real.conds}) have the 
solutions\footnote{Since there is only one representation present we drop
the subscript $u$.} 
\beq
R1:\quad
\begin{array}{|c|c|c|c|}\hline
\tilde c & \tilde \Gamma                            & \tilde \Upsilon                          & \tilde U_1 \cr\hline
+1       & \cos\tilde\theta \mati\otimes\mati       & \sin\tilde\theta \sigma_2\otimes\sigma_2 & \mati      \cr\hline
-1       & \cos\tilde\theta \sigma_3\otimes\sigma_3 & \sin\tilde\theta \sigma_1\otimes\sigma_1 & i \sigma_3 \cr\hline
\end{array}
\qquad
R2:\quad
\begin{array}{|c|c|c|c|} \hline
\tilde c &   \tilde \Gamma                          & \tilde \Upsilon                           & \tilde U_2 \cr\hline
+1       & \cos\tilde\theta \sigma_3\otimes\sigma_3 & i\sin\tilde\theta \sigma_1\otimes\sigma_1 & i\sigma_1  \cr\hline
-1       & \cos\tilde\theta \mati\otimes\mati       & \sin\tilde\theta \sigma_2\otimes\sigma_2  & -\sigma_2   \cr\hline
\end{array}
\label{su2.B.R}
\eeq
where the first matrix in the direct product acts on the flavor indices
and the second on the gauge indices.
We have used the fact that the general solution to $ \tilde U^2 = - \mati $ is
$\tilde U = i \hat\nn \cdot\sigbf $, for an arbitrary (real) unit vector 
$ \hat\nn $; similarly the solutions to $\tilde U \tilde U^* = -\mati $
are $ \tilde U = \pm\sigma_2$, and, finally, the solution to 
$\tilde U \tilde U^* = \mati $ is $ \tilde U = \sin\tilde\alpha
+ i \cos\tilde\alpha \hat\lll\cdot\sigbf $ with $ \tilde\alpha$ real
and $ \hat\lll$ a real unit vector perpendicular to $ \hat\yy $.
We have used the freedom to make global gauge 
rotations to set $ \hat\nn = \hat\zz,~\hat\lll = \hat\xx,~\tilde\alpha=0$.

The general form of $\acal$ follows from (\ref{a.b.unitarity}):
\beq
X_i = u_i W_i \qquad Y_i = v_i W_i^* \sigma_2
\eeq
where $W_i \in$ \su2 and $ |u_i|^2+|v_i|^2 = 1 $. These quantities
are restricted by conditions (\ref{k.l.defs}), (\ref{her.sym}),
(\ref{def.of.s.lam}), (\ref{babaU}), but we will not study
all possible cases as the results are not illuminating. The most important feature of this example
is the presence of non-zero $\Gamma$ {\it and} $\Upsilon$ ($\tilde\Gamma$ {\it and} $\tilde\Upsilon$).

\subsection{$N$ \su3 triplets}

In this section we consider a theory with gauge group \su3
containing $N$ triplets and $\bar N$ anti-triplets. Using the results
obtained at the end of sect. \ref{gen_sol} we can replace all 
anti-triplets by their charge conjugates and obtain a theory where
all fermions transform as a {\boldmath$3$} of \su3; because
of this we take $ \bar N =0 $. In this case
\beq
\tau_a = \left(\begin{array}{cc}
\mati_N \otimes \lambda_a & 0 \cr
0 & - \mati_N \otimes \lambda_a^* 
\end{array} \right)\,,
\eeq
where $\mati_N$ denotes the $ N\times N$ unit matrix in flavor space
and $ \{\lambda_a\}$ denote the usual Gell-Mann matrices\footnote{It is 
worth noting that an \su3 Hermitian matrix can be written in the form
$-(1/3)\mati_3 + \sqrt{4/3} \hat \ell_a \lambda_a $
with $ \sum_a \hat\ell_a^2 = 1~ \hat\ell_a = \sqrt{3} d_{abc} \hat\ell_b \hat\ell_c $
($d_{abc} $ denote the fully-symmetric \su3 symbols).}.

If the boundary condition P1 is chosen then the constraint (\ref{pcon})
implies $ \Gamma = X_1 \otimes U_1^\dagger,~\Upsilon=0 $; in contrast
if P2 is imposed then $ \Gamma =0 , \Upsilon = X_2\otimes U_2^T$
with similar results for $Rj$. These results are summarized in 
the following tables
\beq
\begin{array}{c|cc}
         &  P1 & P2 \cr
\hline
\Gamma   & X_1 \otimes U_1^\dagger & 0                 \cr
\Upsilon & 0                       & X_2 \otimes U_2^T
\end{array}
\qquad
\begin{array}{c|cc}
         &  R1 & R2 \cr
\hline
\tilde\Gamma   & -\tilde X_1 \otimes \tilde U_1^\dagger & 0                 \cr
\tilde\Upsilon &    0                                   & \tilde X_2 \otimes \tilde U_2^T
\end{array}
\eeq
where $X_i,~\tilde X_j $ are unitary $N\times N$ matrices restricted by
the consistency conditions $ \bcal = \bcal^\dagger $ and $ \acal\bcal
= (\acal\bcal)^\dagger $ as specified in (\ref{comp.conds}),
(\ref{k.l.defs}), (\ref{her.sym}) and (\ref{babaU}).

\section{Conditions for the presence of zero KK modes}
\label{lightfer}

In order to extract the 4D particle content of 
this type of theories the standard approach is to expand 
the fields as Fourier series in the compact coordinate. 
For fermion fields the resulting Fourier modes can have
3 possible contributions to their mass: those generated
by $M$, those generated by the $ \bar \Psi \partial_4 \Psi $
term and, finally, those generated by $ \bar \Psi \vevof{A_4} \Psi $
(whenever a non-zero \vev\ is present). The scale of
the last two contributions is set by $1/L$ and is therefore
relatively high. 
The SM light fermions are presumably much lighter than the compactification mass scale $1/L$,
therefore in any realistic setup, the SM fermions are supposed to be zero-modes,
avoiding at least the large $\bar\Psi\partial_4\Psi$ contribution to their mass.
In considering the phenomenology of the class of 5D models discussed in this paper,
it is useful to determine and discuss the general conditions
that allow for the existence of such zero-modes, this is our task for this section.

\def\qwe{\zeta}

Light fermions may exist provided  the boundary conditions allow zero modes
and if $ \vevof{A_4} =0 $ (as can occur if one of the conditions
(\ref{bc.vev}) is not satisfied).
Specifically, we assume that the conditions
$ \chi(y) = \acal^T \chi(y+L) = \gamma_5 \bcal^T \chi(-y) $
allow the expansion of $\chi$ in terms of a complete
set of modes,
$ \chi(x,y) = \sum \chi_n(x) v_n(y) $ 
(examples are provided in section \ref{sect:abelian.case}). 
Massless modes
are associated with a basis function
$v_0$ that is independent of $y$. Writing
$ \chi_0 = (\qwe^c,\qwe)^T$ and substituting 
into (\ref{period}) and (\ref{Rbc}) gives
\beq
(\mati-\acal^*) \chi_0 =0 \qquad (\mati - \gamma_5 \bcal^*) \chi_0 =0 \,.
\label{f0m}
\eeq
In order to avoid having $\chi_0 = 0$ as the only solution we must have 
det$(\mati-\acal^*)=$det$(\mati- \gamma_5 \bcal^*)=0$.

Using (\ref{adef}) and (\ref{bdef}) these constraints on the light modes become
\beq
\begin{array}{ll}
\left(\mati -\Gamma \right) \qwe_L = \Upsilon^* \left(\qwe_R \right)^c & \qquad
\left(\mati +\tilde\Gamma \right) \qwe_L = 
- \tilde\Upsilon^* \left(\qwe_R \right)^c \cr
\Upsilon  \qwe_L = \left(\Gamma^*-\mati \right)\left(\qwe_R \right)^c & \qquad
\tilde\Upsilon  \qwe_L = \left(\tilde\Gamma^*-\mati \right)
\left(\qwe_R \right)^c \,.
\end{array}
\label{light}
\eeq
Assuming $\Upsilon=\tilde\Upsilon=0$, the only possible solution 
of the above equations
is $ \Gamma=\mati$ and $ \tilde\Gamma = \mati$ with $ \qwe_L =0 $ or $ \Gamma=\mati$ 
and $ \tilde\Gamma = -\mati $ with $\qwe_R =0 $. It is worth noting that
this special case is the standard strategy adopted in the context of universal extra dimensions in order
to construct chiral effective 4D theory.

Other solutions must be considered on a case by case basis. Note however that
if det$\tilde\Upsilon \not=0 $, then the two equations
involving $ \tilde\Gamma$ and $\tilde \Upsilon $ are equivalent:
from $ \bcal = \bcal^\dagger = \bcal^{-1} $ we find
$ \tilde\Upsilon^* \tilde\Upsilon = \mati - \tilde\Gamma^2 $
and $ \tilde\Upsilon \tilde\Gamma = \tilde\Gamma^* \tilde \Upsilon$,
so, if $ \tilde \Upsilon$ has an inverse, so do
$ \mati \pm\tilde\Gamma $; this also shows that
$ \tilde\Upsilon^*{}^{-1} (\mati + \tilde\Gamma)  = 
(\mati - \tilde\Gamma^*)^{-1} \tilde\Gamma$
which proves the assertion. Therefore one of the constraints involving 
$ \tilde\Gamma$ and $\tilde \Upsilon $ in ({\ref{light}) can be dropped. 

From (\ref{Pbc}) and (\ref{Rbc}) one can easily find the necessary conditions 
which must be fulfilled for gauge boson zero modes to exist. Denoting by
$\hat a,~\hat b,\ldots$ the gauge indices associated with these zero modes
we find that the 4D gauge fields $A^\mu_{\hat a}$ have a zero mode provided
$ [ \tau^{\hat{a}}, \ucal_i] = 0$ and $ [\tau^{\hat{a}}, \tilde\ucal_j] =0 $. 
The zero mode of $A^4_a$ is present if $ [ \tau^{\hat{a}}, \ucal_i] = 0$ 
and $ \{\tau^{\hat{a}}, \tilde\ucal_j\} =0 $. 

As we have already observed in section~\ref{sect:abelian.case},
the fermionic zero modes may satisfy the generalized Majorana 4D condition:
\beq
\zeta = N C_4 (\bar\zeta)^T\,,
\label{Mc}
\eeq
where $C_4$ is the 4D charge conjugation operator\footnote{In the Dirac representation
$C_4=\gamma_0\gamma_2$ while the 5D one is $C_5=\gamma_1\gamma_3$. It is useful to note
that $\gamma_5C_5=-iC_4$.}, and $ N$ acts on flavor and gauge indices. In this case
we can express $\zeta$ as
\beq
\zeta = \left( \begin{array}{c} N \sigma_2 \varphi^* \cr \varphi \end{array} \right)\,,
\eeq
where $ \varphi$  denotes a 2-component spinor and $ \sigma_2 $ acts on the 
Lorentz indices. Consistency of this expression requires $N N^* = \mati $.

For the Majorana spinor $\zeta$, the conditions (\ref{f0m}) become
\beq
\baa{lll}
(\mati - \Gamma) \varphi + i \Upsilon^* \sigma_2 \varphi^* =0\,, &
(N^*\Gamma-\Gamma^*N^*)\varphi=0\,, & (N^*\Upsilon^*-\Upsilon N)\varphi^*=0\\
(N + i\tilde\Upsilon^*)\sigma_2\varphi^* - \tilde\Gamma\varphi=0\,, &
(N^*\tilde\Gamma + \tilde\Gamma^*N^*)\varphi=0\,, &
(N^*\tilde\Upsilon^* + \tilde\Upsilon N)\varphi^* =0\,.
\eaa
\label{mfer}
\eeq

It is useful to illustrate the above conditions by certain special cases:
\bit
\item
If $ \Upsilon  = 0$ and $\Gamma=\mati$ ($\acal=\mati$, so periodic fermionic fields), and  $\tilde\Upsilon =0 $ then 
it is easy to see from (\ref{mfer}) that more than one flavour is needed to have a Majorana zero mode.
\item
If $\tilde\Gamma=0$~\footnote{The necessary existence of $\tilde\Upsilon^{-1}$ is guaranteed by the unitarity
of $\bcal$. In this case a charge conjugated field appears in the orbifold BC.} 
then there is always a Majorana zero mode with $N=-i\tilde\Upsilon^\dagger$. 
This case is illustrated by the BC $(P1-R2)$ for a single Abelian fermion, if
$ u=1,~v=0 $ ($ \alpha =0 $) are chosen, see  (\ref{one.f.p.r}), then $N=-i$.
\item
If $\tilde\Upsilon=0$~\footnote{Again the unitarity of $\bcal$ shows that $\tilde\Gamma\neq 0$.}, $\Gamma^*\neq \mati$
and $\Upsilon$ is invertible (so charge conjugated field appears in the periodicity BC) 
then again there exists a Majorana zero mode if  $N=-i\Upsilon^{-1} (\mati - \Gamma^*) \tilde\Gamma^*$
and if this matrix satisfies the constraints of the last two columns in (\ref{mfer}).
For an Abelian model this again requires more than one flavor: the single fermion case would 
correspond to the  $(P2-R1)$ BC for which, using (\ref{one.f.p.r}), 
$ \tilde\Gamma=s_b,~\Upsilon=1,~\Gamma=\tilde\Upsilon=0 $. In this case, however
$N^*\tilde\Gamma + \tilde\Gamma^* N^* = 2i$, so that the corresponding equation in (\ref{mfer})
implies $\varphi =0 $.
\eit

\subsection{Examples}
\label{lightfer.ex}

A simple situation that allows for the presence of light modes is
realized by taking $ \acal = \mati $ (periodic fermions), and assuming 
$ \det\tilde\Upsilon \not=0 $,
then we have
\beq
( \zeta_R)^c = - (\tilde\Upsilon^*)^{-1} ( \mati+ \tilde\Gamma) \zeta_L \,,
\label{lin_rel}
\eeq 
or, equivalently,
\beq
\qwe = \left( \begin{array}{c}
\tilde\Gamma\phi - \tilde\Upsilon^* i \sigma_2\phi^* \cr \phi 
\end{array}\right)\,,
\eeq
where $\phi$ is a two-component spinor ($\sigma_2$ acts
on Lorentz indices, $\tilde\Gamma$ and $ \tilde\Upsilon$
on flavor indices). 
Using unitarity of $\bcal$ and invariance of the mass term under the orbifold twist
one can show that in this case the
mass term in the Lagrangian becomes
\beq
\bar\qwe M \qwe = - 2 \phi^\dagger M \phi 
\eeq
having taken $M$ real and diagonal. The diagonal elements of $M$ could be chosen arbitrarily small.
Note that if $\tilde\Gamma=0$ then the zero mode $\zeta$
is a Majorana type fermion. 

For a specific example let us consider
an \su2 theory containing two doublets
(we still assume $ \acal = \mati $) . The representation
is pseudoreal generated by the Pauli matrices $ \sigma_I $; $ \sigma_2 $
plays the role of the matrix $S_u$ of section \ref{gen_sol}. In this
case we will write $ \zeta^T  = (\psi_1, \psi_2)^T $ where $ \psi_i $
($ i = 1,2 $) are doublets with the flavor index $i$ (the gauge index is not
displayed). The matrix $ \bcal$ was determined in section \ref{su2}, see (\ref{su2.B.R}).

The simplest case corresponds to $ \tilde c=1$ in (\ref{su2.B.R}), which we assume.
This case illustrates the interesting possibility
of non-zero $\tilde\Gamma$ and $\tilde\Upsilon$, which can occur only if there exists
at least two multiplets transforming according to the same pseudo-real
representation, as was mentioned at the end of section~\ref{gen_sol}.
We find that (\ref{su2.B.R}) and (\ref{light})
imply $ P_R\psi_{1,2} = \mp i \cot(\tilde\theta/2) \sigma_2 (P_L \psi_{2,1})^c $
(where $ \sigma_2 $ acts on the gauge indices). In particular $ \psi_2 $
can be expressed in terms of $ \psi_1$:
\beq
\psi_2 = i\left(\frac{\cos\tilde\theta + \gamma_5}{\sin\tilde\theta} \right) 
\sigma_2 \psi_1^c \; \qquad (R1,~\tilde c=+1)\,.
\eeq

The light modes $\zeta$ can acquire a small mass of order $M$
provided $ \{\bcal,\mcal\} = 0 $. This can occur for R1 and $ \tilde c=-1$
or R2 and $ \tilde c = +1 $: in either of these cases $M$ should satisfy 
$\{\sigma_3 ,M\} =0 $ and $ \sigma_1 M \sigma_1 = M^* $, so that
$ M = m_1 \sigma_1 + m_2 \sigma_2,~~m_{1,2}$ real;
the physical masses are simply $ \pm \sqrt{m_1^2 + m_2^2}$.

 Abelian examples of zero modes were briefly mentioned in
section \ref{sect:abelian.case}. The possibility that the gauge fields $A^4$  
acquire a \vev\ contributing to the fermion mass will
not be discussed in detail here.

\subsection{Gauge invariance}
\label{lightfer.gi}

The conditions (\ref{light}) need not be invariant under arbitrary $y$-independent 
gauge transformations $ \Omega $, $ \zeta \to \Omega(x) \zeta $
(see (\ref{gauge.transf})), leading to a reduction of the gauge group for the 
light sector. The specific constraints follow from the
decomposition (\ref{form.of.Ta}) that allows us to write
$ \Omega = \diag(\cdots, \mati_{d_\ell}\otimes\Omega_\ell , \cdots \mati_{d_u}\otimes\Omega_u , \cdots)$
where $ \Omega_r$ denotes a gauge transformation in the space
corresponding to the \irrep\ $r$.

Using this we then find from (\ref{light}) (or equivalently from (\ref{gauge.inv}) for $y$-independent $\Omega$)
that for non-complex representations the BC are preserved by gauge transformations that obey
\beq
\begin{array}{lcl}
P1:~[U_{1;u},\Omega_u] =0               & \qquad & R1:~[\tilde U_{1;u},\Omega_u] =0 \cr
P2:~[S_u^\dagger \cdot U_{2;u},\Omega_u] =0     & \qquad & R2:~[S_u^\dagger  \cdot \tilde U_{2:u},\Omega_u] =0\,,
\end{array}
\label{light_real.gi}
\eeq
where we use the same notation as in sec. \ref{gen_sol} 
(see \cite{Hebecker:2001jb} for related arguments in the case of standard BC).

It is easy to see that 
the $ \Omega $ satisfying the constraints (\ref{light_real.gi}) form a subgroup 
of the original gauge group and, in general that its representation 
is complex. It is therefore possible to use BC to 
select a light sector that has both a chiral structure and a 
smaller gauge group. Such a breaking has clear value in model building
and will be investigated in a future publication.

For complex representations we  find the same conditions
for P1 and R1, but not when P2 or R2 are imposed:
\beq
\begin{array}{lcl}
P1:~[U_{1;u},\Omega_\ell] =0               & \qquad & R1:~[\tilde U_{1;u},\Omega_\ell] =0 \cr
P2:~\Omega_\ell^* = U_{2;\ell} \Omega_\ell U_{2;\ell}^\dagger     & \qquad & 
R2:~\Omega_\ell^* =\tilde U_{2;\ell} \Omega_\ell \tilde U_{2;\ell}^\dagger\,.
\end{array}
\label{light_comp.gi}
\eeq
For P2 and R2 these conditions cannot be satisfied by all elements of the
initial gauge group (since they would then imply that the representation is non-complex), 
so the gauge group of the light sector must be
smaller then the initial group, and the light sector fermions will transform according to a non-complex
representation of this subgroup\footnote{
An example would be an \su3 theory with fermions in the fundamental 
(complex) representation where the light sector gauge group is reduced
to \su2 that has only non-complex representations.}.

Let us now consider the gauge invariance of the gauge boson light sector.
As it was already mentioned earlier, the  gauge fields $A^\mu_{\hat{a}}$ will have a zero mode provided
\beq
\begin{array}{lcl}
P1:~+T_{\hat{a}}=U_1 T_{\hat{a}} U_1^\dagger     & \qquad & R1:~+T_{\hat{a}}=\tilde U_1 T_{\hat{a}} \tilde U_1^\dagger    \cr
P2:~-T_{\hat{a}}^*=U_2T_{\hat{a}}U_2^\dagger     & \qquad & R2:~-T_{\hat{a}}^*=\tilde U_2T_{\hat{a}}\tilde U_2^\dagger\,.
\end{array}
\label{light.b}
\eeq
Comparing (\ref{light_real.gi}) and (\ref{light_comp.gi}) with (\ref{light.b}), it is easy to see that 
both for non-complex and for complex representations, the generators $T_{\hat{a}}$ that correspond to 
zero modes generate the symmetry group of the light sector. In other words, the gauge symmetry
of the zero mode sector can be easily determined just by inspection
of the massless vector bosons.

As an example let us consider here \su2 gauge theory with a single doublet of fermions.
We adopt the $P1-R1$ BC and choose
\beq
\begin{array}{lll}
\Gamma=\sigma_3         & \Upsilon =0           & U_1 = i \sigma_3 \cr
\tilde\Gamma=\mati      & \tilde\Upsilon =0     & \tilde U_1 = \mati\,,
\end{array}
\eeq
which is a slight modification of the example discussed at the end of section~\ref{sect:gi}.
In that section we found that the symmetry of the light sector was completely broken due to
non-trivial orbifold BC ($\bcal \neq \mati$). Here we choose $\bcal = \mati$ so that the
$y$-independent gauge transformations generated by $ \sigma_3 $ are allowed, as a 
consequence the zero mode of $A_3^\mu$ survives. In turn, the light fermion modes obey 
\beq 
\zeta_L =0 \qquad \left( \sigma_3 - \mati \right) \left(\zeta_R\right)^c =0 \,.
\eeq
The light-sector contains only the $A_3^\mu$ massless gauge bosons 
and a right-handed, charged (and therefore massless), fermion.

\section{Summary and Conclusions}
\label{summary}

In this paper we considered a generic gauge theory in a  5-dimensional
space compactified on $ \mBB^4 \times(S_1/Z_2)$, and studied the
effects of a generalized set of boundary conditions (BC) that allow for 
mixing between  particles and anti-particles after a translation by the 
size of the extra dimension or after an orbifold reflection. 

We described the consequences of gauge invariance as well as the general
form of the boundary conditions consistent with it. We also studied
the behavior of this class of theories under 5D parity(P), charge 
conjugation(C) and CP. In particular we determined the conditions under
CP will be violated (explicitly) by the BC
as well as spontaneously, through a possible \vev\ of the fifth component
of the gauge fields.

We derived a simple set of conditions that determine the light-particle
content of the model and the corresponding gauge subgroup, noting also
the possibility that the light fermions might have  chiral structure
and transform under complex representations of the light-sector gauge group,
even though the underlying theory is vector-like and contains only real
representations of the full gauge group. In addition we derive the conditions under
which the  model generates light Majorana particles.

We believe these aspects will be of relevance in constructing phenomenologically
acceptable theories.

The general considerations were illustrated by many Abelian and non-Abelian examples.

\appendix
\section{One Abelian fermion}

Here we will consider the one fermion case with $(P1-R2)$ BC.
The fermion mass term (\ref{p1-r21fer}), 
can be brought to the standard real form $m\up{\rm phys}_n \bar\psi_n\psi_n$ 
through the following chiral rotation\footnote{
The chiral rotation of the fermions induces an 
$\epsilon_{\mu\nu\sigma\rho}F_{\mu\nu}F^{\sigma\rho}$ 
term in the Lagrangian;
however, in the Abelian case considered here, this is a total 
derivative and therefore it can be dropped.}: 
\beq
\psi_n \to \exp( i \gamma_5 \theta_n) \psi_n \quad {\rm with} \quad 
\tan( 2 \theta_n) = { 2\pi n +\alpha \over  M L}; \quad
| \theta_n| \le \pi/4
\label{rot.fer}
\eeq
From this we find that the physical fermion masses\footnote{In order
to include contributions from non-zero \vev\ of $A_4$ one should 
replace $\alpha$ by $( \alpha - g_5 q L^{1/2} \langle A^4_0\rangle )$.} are
\beq
m\up{\rm phys}_n = \sqrt{ M^2 + \left({2\pi n +\alpha \over L}\right)^2}\,,
\eeq 

Form the orbifold conditions and from the 
reality of the gauge field 
we have the following constraints for the bosonic KK modes:
\bea
A^\mu_n &=& -A^\mu_{-n} = \left(A^\mu_{-n}\right)^* \cr
  A^4_n &=& + A^4_{-n} =  \left(A^4_{-n}  \right)^*
\eea
The above conditions allow to rewrite the Lagrangian in terms of 
non-negative modes only. In addition
one can see that $A^\mu_{n=0} = \re A^\mu_n = \im A^4_n = 0$. 
Adding the gauge fixing Lagrangian,
\beq
\lcal_{gf} = -\inv{2\xi} \left( \partial_\mu A^\mu + 
\xi \partial_4 A^4 \right)^2
\eeq
the gauge kinetic energy terms read
\bea
\lcal_A + \lcal_{gf} &=&-\frac12\scalar \,\square\,\scalar + 
\frac12\sum_{n=1}^\infty\left\{
B_n^\mu\left[(\square+\omega_n^2) g_{\mu\nu}
-(1-\xi^{-1}\partial_\mu\partial_\nu)\right]B_n^\nu
-B^4_n (\square + \xi\omega_n^2) B^4_n \right\}\,,\non
\eea
where $ \lcal_A = - (F_{M N})^2/4 $ and where we defined the fields
\beq
\scalar \equiv - \sqrt{L} \re A^4_{n=0} , \qquad
B^\mu_n \equiv \sqrt{2L} \im A^\mu_n , \qquad
B^4_n \equiv \sqrt{2 L} \re  A^4_n\,.
\eeq

In terms of the $B^N$ and the rotated fermion fields 
defined in (\ref{rot.fer}) we obtain 
\bea
\int_0^L dy \bar\Psi ( i \gamma^N D_N -M ) \Psi
&=& \sum_n\bar\psi_n \left[i \not\!\partial - m_n\up{\rm phys} \right]\psi_n
%
%
- g \scalar \sum_n \bar\psi_n \left[\sin(2 \theta_n) -
 i \gamma_5  \cos(2\theta_n)  \right]\psi_n 
\cr && 
- \frac g{\sqrt{2}}  \Biggl\{\hspace{3pt}
i \sum_{k>l} \bar\psi_k \left[ \cos(\theta_k-\theta_l) +
i \gamma_5 \sin(\theta_k-\theta_l)  \right]\not \!\! B_{k-l} \psi_l 
\cr && \hspace{30pt} 
+\sum_{k>l} B^4_{k-l} \bar\psi_k \left[ 
\sin(\theta_k+\theta_l) - i \gamma_5 \cos(\theta_k+\theta_l)\right]\psi_l + 
\hbox{H.c.} \Biggr\}\,,
\eea
where $ g = g_5/\sqrt{L}$

From these expressions one can obtain the effective potential
for the scalar $ \scalar$. Following \cite{Grzadkowski:2004jv}
(including an additional factor of $1/2$ since we are dealing with
Majorana fermions) we find
\beq
V_{\rm eff} = \inv{4 \pi^2 L^4} \re \left[
{\rm Li}_5(\zeta) + 3 x {\rm Li}_4(\zeta) + x^2 {\rm Li}_3(\zeta) 
\right]\,,
\eeq
where 
\beq
x = M L \qquad \zeta = e^{ - x + i\left(\alpha - g L \vevof\scalar\right)}\,.
\eeq

The absolute minima occur at $ \alpha - g L \vevof\scalar = 
\pm \pi,\pm3 \pi,\ldots $, but this does not lead to any physical
CP violation effects unless we add a second fermion with
different charge and/or different $ \alpha$ (modulo $\pi $), see \cite{Grzadkowski:2004jv}.

\vspace*{0.5cm}
\centerline{\bf ACKNOWLEDGMENTS}
\vspace*{0.5cm}
B.G. thanks Zygmunt Lalak, Claudio Scrucca, Fabio Zwirner for discussions.
J.W. thanks F. del \'Aguila for illuminating comments.
This work is supported in part by the State
Committee for Scientific Research (Poland) under grant 1~P03B~078~26,
and by funds provided by the U.S. Department of Energy under grant No.
DE-FG03-94ER40837.


\begin{thebibliography}{99}
\bibitem{Georgi:1974yf}
H.~Georgi, H.~R.~Quinn and S.~Weinberg,
Phys.\ Rev.\ Lett.\  {\bf 33}, 451 (1974).
%
\bibitem{Cremmer:1979uq}
E.~Cremmer, J.~Scherk and J.~H.~Schwarz,
``Spontaneously Broken N=8 Supergravity,''
Phys.\ Lett.\ B {\bf 84}, 83 (1979);\\
%
J.~Scherk and J.~H.~Schwarz,
``How To Get Masses From Extra Dimensions,''
Nucl.\ Phys.\ B {\bf 153}, 61 (1979).

\bibitem{Hosotani:1983xw}
Y.~Hosotani,
``Dynamical Mass Generation By Compact Extra Dimensions,''
Phys.\ Lett.\ B {\bf 126}, 309 (1983);\\
%
Y.~Hosotani,
``Dynamical Gauge Symmetry Breaking As The Casimir Effect,''
Phys.\ Lett.\ B {\bf 129}, 193 (1983);\\
%
A.~T.~Davies and A.~McLachlan,
``Gauge Group Breaking By Wilson Loops,''
Phys.\ Lett.\ B {\bf 200}, 305 (1988);\\
%
Y.~Hosotani,
``Dynamics Of Nonintegrable Phases And Gauge Symmetry Breaking,''
Annals Phys.\  {\bf 190}, 233 (1989);\\
%
A.~T.~Davies and A.~McLachlan,
``Congruency Class Effects In The Hosotani Model,''
Nucl.\ Phys.\ B {\bf 317}, 237 (1989);\\
%
H.~Hatanaka, T.~Inami and C.~S.~Lim,
``The gauge hierarchy problem and higher dimensional gauge theories,''
Mod.\ Phys.\ Lett.\ A {\bf 13}, 2601 (1998)
[arXiv:hep-th/9805067];\\
%
H.~Hatanaka,
``Matter representations and gauge symmetry breaking via compactified
space,''
Prog.\ Theor.\ Phys.\  {\bf 102}, 407 (1999)
[arXiv:hep-th/9905100];\\
%
Y.~Kawamura,
``Gauge symmetry reduction from the extra space S(1)/Z(2),''
Prog.\ Theor.\ Phys.\  {\bf 103}, 613 (2000)
[arXiv:hep-ph/9902423];\\
%
L.~J.~Hall, H.~Murayama and Y.~Nomura,
``Wilson lines and symmetry breaking on orbifolds,''
Nucl.\ Phys.\ B {\bf 645}, 85 (2002)
[arXiv:hep-th/0107245];\\
%
M.~Kubo, C.~S.~Lim and H.~Yamashita,
``The Hosotani mechanism in bulk gauge theories with an orbifold extra  space
S(1)/Z(2),''
Mod.\ Phys.\ Lett.\ A {\bf 17}, 2249 (2002)
[arXiv:hep-ph/0111327];\\
%
N.~Haba, M.~Harada, Y.~Hosotani and Y.~Kawamura,
``Dynamical rearrangement of gauge symmetry on the orbifold S(1)/Z(2),''
Nucl.\ Phys.\ B {\bf 657}, 169 (2003)
[Erratum-ibid.\ B {\bf 669}, 381 (2003)]
[arXiv:hep-ph/0212035];\\
%
K.~Takenaga,
``Effect of bare mass on the Hosotani mechanism,''
Phys.\ Lett.\ B {\bf 570}, 244 (2003)
[arXiv:hep-th/0305251];\\
%
N.~Haba, Y.~Hosotani and Y.~Kawamura,
``Classification and dynamics of equivalence classes in SU(N) gauge theory on
the orbifold S(1)/Z(2),''
Prog.\ Theor.\ Phys.\  {\bf 111}, 265 (2004)
[arXiv:hep-ph/0309088];\\
%
N.~Haba, Y.~Hosotani, Y.~Kawamura and T.~Yamashita,
``Dynamical symmetry breaking in gauge-Higgs unification on orbifold,''
arXiv:hep-ph/0401183;\\
%
N.~Haba and T.~Yamashita,
``The general formula of the effective potential in 5D SU(N) gauge theory on
orbifold,''
JHEP {\bf 0402}, 059 (2004)
[arXiv:hep-ph/0401185].

\bibitem{Quiros:2003gg}
M.~Quiros,
``New ideas in symmetry breaking,''
arXiv:hep-ph/0302189.

\bibitem{Csaki:2003dt}
C.~Csaki, C.~Grojean, H.~Murayama, L.~Pilo and J.~Terning,
``Gauge theories on an interval: Unitarity without a Higgs,''
Phys.\ Rev.\ D {\bf 69}, 055006 (2004)
[arXiv:hep-ph/0305237];\\
%
C.~Csaki, C.~Grojean, L.~Pilo and J.~Terning,
``Towards a realistic model of Higgsless electroweak symmetry breaking,''
Phys.\ Rev.\ Lett.\  {\bf 92}, 101802 (2004)
[arXiv:hep-ph/0308038].

\bibitem{SekharChivukula:2001hz}
R.~Sekhar Chivukula, D.~A.~Dicus and H.~J.~He,
``Unitarity of compactified five dimensional Yang-Mills theory,''
Phys.\ Lett.\ B {\bf 525}, 175 (2002)
[arXiv:hep-ph/0111016];\\
%
R.~S.~Chivukula, D.~A.~Dicus, H.~J.~He and S.~Nandi,
``Unitarity of the higher dimensional standard model,''
Phys.\ Lett.\ B {\bf 562}, 109 (2003)
[arXiv:hep-ph/0302263];\\
%
S.~De Curtis, D.~Dominici and J.~R.~Pelaez,
``Strong tree level unitarity violations in the extra dimensional standard
Phys.\ Rev.\ D {\bf 67}, 076010 (2003)
[arXiv:hep-ph/0301059];\\
%
Y.~Abe, N.~Haba, K.~Hayakawa, Y.~Matsumoto, M.~Matsunaga and K.~Miyachi,
``4D equivalence theorem and gauge symmetry on orbifold,''
arXiv:hep-th/0402146.

\bibitem{Grzadkowski:2004jv}
B.~Grzadkowski and J.~Wudka,
Phys.\ Rev.\ Lett.\  {\bf 93}, 211603 (2004)
[arXiv:hep-ph/0401232].

\bibitem{BelenGavela:1983tp}
M.~Belen Gavela and R.~I.~Nepomechie,
``Discrete Symmetries In Kaluza-Klein Theories,''
Class.\ Quant.\ Grav.\  {\bf 1}, L21 (1984).
%
K.~Shimizu,
``C, P And T Transformations In Higher Dimensions,''
Prog.\ Theor.\ Phys.\  {\bf 74}, 610 (1985).

\bibitem{Biggio:2002rb}
C.~Biggio, F.~Feruglio, A.~Wulzer and F.~Zwirner,
``Equivalent effective Lagrangians for Scherk-Schwarz compactifications,''
JHEP {\bf 0211}, 013 (2002)
[arXiv:hep-th/0209046].


\bibitem{Scrucca:2003ra}
C.~A.~Scrucca, M.~Serone and L.~Silvestrini,
``Electroweak symmetry breaking and fermion masses from extra dimensions,''
Nucl.\ Phys.\ B {\bf 669}, 128 (2003)
[arXiv:hep-ph/0304220].

\bibitem{Lee:1966ik}
T.~D.~Lee and G.~C.~Wick,
``Space Inversion, Time Reversal, And Other Discrete Symmetries In Local Field
Phys.\ Rev.\  {\bf 148}, 1385 (1966).

\bibitem{Hebecker:2001jb}
A.~Hebecker and J.~March-Russell,
``The structure of GUT breaking by orbifolding,''
Nucl.\ Phys.\ B {\bf 625}, 128 (2002)
[arXiv:hep-ph/0107039].


\end{thebibliography}
\end{document}